\documentstyle[11pt]{article}
\def\appendix{\par
 \setcounter{section}{0}
 \setcounter{subsection}{0}
 \def\thesection{Appendix \Alph{section}}
 \def\theequation{\Alph{section}.\arabic{equation}}
 \setcounter{equation}{0}}

\begin{document}
\begin{flushright}
(revised version of hep-ph/9501268, DO-TH-94/20, THES-TP-94/06)\\
 September 1995
\end{flushright}
\centerline{\Large \bf Condensation effects beyond one loop in the}
\centerline{\Large \bf Top-mode Standard Model without gauge bosons}
\vspace{1.2cm}
\begin{center}
{{\bf G.~Cveti\v c} and {\bf E.A.~Paschos} \\
Inst.~f\"ur Physik, Universit\"at Dortmund, 44221 Dortmund, Germany\\
[0.8cm]
{\bf N.D.~Vlachos}\\
Dept.~of Theor.~Physics, Aristotle University of Thessaloniki,
540 06 Thessaloniki, Greece }
\end{center}

\vspace{1.2cm}
\centerline{\bf Abstract}
We study dynamical symmetry breaking in the Standard Model including the
next-to-leading order terms. We introduce at a high, but finite, energy scale
$\Lambda$ a top quark condensate $H=\langle t \bar t \rangle$ and derive,
using path integral methods, the effective potential including quadratic
fluctuations in the scalar field $H$. We neglect the contributions of all
components of the massive electroweak gauge bosons. The existence of a
non-trivial minimum in the effective potential leads to the condition that
the cut-off $\Lambda$ is limited from above: $\Lambda \stackrel{<}{\approx}
\Lambda_{\mbox{\footnotesize crit}} \approx 4.7 m_t^{\mbox{\footnotesize phys.
}}$ (for $N_{\mbox{\footnotesize c}}=3$), and to a new lower bound for the
4-fermion coupling $a =(G N_{\mbox{\footnotesize c}} \Lambda^2)/(8 \pi^2)
\stackrel{>}{\approx} 1.60$. Similar results are obtained if we demand,
instead, that the next-to-leading order contributions not shift the location
$z=(m_{\mbox {\footnotesize t}}^{\mbox {\footnotesize bare}}/\Lambda)^2$ of
the minimum drastically, e.g.~by not more than a factor of 2. The results are
reproduced diagrammatically, where the leading plus all the next-to-leading
order diagrams in the $1/N_{\mbox{\footnotesize c}}$-expansion are included.
Dominant QCD effects are also included, but their impact on the numerical
results is shown to be small.\\
PACS number(s): 11.15.Pg, 12.50.Lr


\section{Introduction}

Inspired by the pioneering work of Nambu and Jona-Lasinio (NJL)~\cite{njl},
many studies appeared in the past which consider the Higgs
mesons as bound states (condensates) of heavy quark pairs
(~\cite{n}-\cite{suzuki} and references therein) . Usually, the 1-loop
effective potential of the Higgs condensate is calculated starting with
effective 4-fermion interactions, whose origin is a new as yet unknown
physics at higher energies. While the Higgs field is assumed to be
auxiliary (non-dynamical) at the outset, it becomes dynamical
through (1-loop) quantum effects. In order to investigate qualitative
features of condensation and to avoid additional technical difficulties,
the calculation is frequently performed by ignoring the effects of
the electroweak gauge bosons and taking into account only the
effects of fermionic loops - this is frequently called the
large-$N_{\mbox{\footnotesize c}}$
approximation. The vacuum expectation value (VEV) is the value of the
Higgs field at the minimum of the effective potential.
This condition is a relation between the
mass of the heavy quark (usually the top quark), the 4-fermion
coupling parameters and the energy cut-off $\Lambda$ where the
condensation occurs. The relation is called the 1-loop
gap equation.

The dynamical symmetry breaking (DSB), as manifested by the condensation,
is essentially a non-perturbative phenomenon. Nonetheless, the effective
potential can be calculated perturbatively. At 1-loop level, the summation
of an infinite number of the relevant diagrams, for any fixed value of the
internal 4-momentum $\bar k$, leads to an analytic function which gives
naturally the analytic continuation into regions of low energy
(${\bar k}^2 < m_{t} ^2$) where the original perturbative series does not
converge~\cite{gceap}. These analytic functions can be subsequently
integrated over the internal momenta $\bar k$, incorporating in this way
the non-perturbative regions.

Here we emphasize that we consider the effective potential as a
function of only a hard mass term $\lambda \sigma_0$ of the top quark,
parametrized by the expectation value $\sigma_0$ of a composite
(initially auxiliary) scalar field $\sigma$. On the other hand,
we note that there exists a more general formalism for the DSB
developed by Cornwall, Jackiw and Tomboulis~\cite{cjt}, which is
variational and studies the general functional
form $\Sigma ( {p \llap /} )$ of the expectation value of the
proper self-energy part in the quark propagator.

In this article, we derive the effective potential $V_{\mbox
{\footnotesize eff}}$ beyond the one loop by the path integral method, and
later on we reproduce the results diagrammatically, summing up the leading
plus the next-to-leading order diagrams in the
$1/N_{\mbox{\footnotesize c}}$-expansion. We work within the framework of
the NJL-inspired Top-mode Standard Model (TSM)~\cite{bhl} and neglect the
effects of all components of the electroweak (massive) gauge bosons.
Although path integral methods are to a certain extent known in the
literature of the NJL-type models (cf.~\cite{higa} and references therein),
we decided, for reasons of clarity, to present in Section 2 a comprehensive
derivation of the effective potential $V_{\mbox{\footnotesize eff}}$ using
these methods. We include there the contribution of
$V^{\mbox{\scriptsize (ntl)}}_{\mbox{\footnotesize eff}}$ produced
by quadratic quantum
fluctuations of the Higgs condensate around its ``classical'' value $H_0$.
Technical details of this calculation are given
in Appendices A and B. In Appendix C, we rederive
$V^{\mbox{\scriptsize (ntl)}}_{\mbox{\footnotesize eff}}$
diagrammatically and demonstrate that it represents the
contributions of all those $(\ell + 1)$-loop diagrams
($\ell=1,2,3,\ldots.$) which are next-to-leading order (${\cal{O}} (1)$)
in the formal $1/N_{\mbox{\footnotesize c}}$-expansion\footnote{ The
superscript ``ntl'' stands
throughout the paper for the {\it next-to-leading} order in the
$1/N_{\mbox{\footnotesize c}}$-expansion.
The $V^{(1)}_{\mbox{\footnotesize eff}} $, originating from the 1-loop
1-PI diagrams, represents at the same time all the {\it leading} order
(${\cal {O}}(N_{\mbox{\scriptsize c}})$) contributions to
$V_{\mbox{\footnotesize eff}}$
in the $1/N_{\mbox{\footnotesize c}}$-expansion.}.
We then calculate in Section 3 the dominant part of the QCD contributions to
$V_{\mbox{\footnotesize eff}}$ using the diagrammatic method and demonstrate
that their numerical impact is small. In Section 4 we investigate the minimum
of $V_{\mbox{\footnotesize eff}}$ and arrive at an improved gap equation --
an improved relation between $m_{t} $, $\Lambda$ and the 4-fermion coupling
parameter $G$. The requirement of the existence of a non-trivial minimum
(non-zero VEV) in $V_{\mbox{\footnotesize eff}}$ leads to an additional
restriction (upper bound) on the value of the physically interesting
parameter $\Lambda/m_{t} $. We demonstrate explicitly that
the existence of a non-trivial minimum of the effective potential,
within the present framework, implies an upper bound for $\Lambda/m_{t} $
of ${\cal {O}}(10^1)$. Similar results are obtained when the requirement of
the existence of the minimum is replaced by the requirement that the
next-to-leading order contributions, discussed in this paper, not
shift the location of the minimum drastically. Consequently,
the energy $\Lambda$, where the top and antitop condense to form
the Higgs, cannot be very high. Within the present framework we discuss
in Section 5 renormalization corrections to the mass $m_t$.
The new result of the present analysis is an upper bound for $\Lambda$
of the order of TeV, which is absent in previous studies of Top-mode
Standard Model using the 1-loop gap equation.

\section{Calculation of the effective potential beyond one loop -
 contributions of the scalar sector}

The Top-mode Standard Model (TSM)~\cite{bhl} is a framework with a truncated
gauge-invariant 4-fermion interaction at a high energy scale $E \sim
\Lambda$
\begin{equation}
{\cal{L}} = {\cal{L}}^0_{\mbox {\footnotesize kin}}
 + G \left( \bar \Psi^{ia}_{\mbox{\scriptsize L}}
 t_{\mbox{\scriptsize R} a} \right)
\left( \bar t^b_{\mbox{\scriptsize R}}
\Psi^i_{\mbox{\scriptsize L}b} \right)
 \qquad \mbox{for} \ E \sim \Lambda \ ,
\label{TSM}
\end{equation}
where the color indices
($a$,$b$) are summed over, $\Psi^{T}_{\mbox{\scriptsize L}}
= (t_{\mbox{\scriptsize L}},
b_{\mbox{\scriptsize L}})$, and ${\cal{L}}^0_{\mbox {\footnotesize kin}}$
represents the familiar gauge-invariant kinetic terms of fermions and
gauge bosons. The Lagrangian (\ref{TSM})
can be rewritten in terms of an as yet auxiliary scalar field $H$
\begin{equation}
{\cal{L}} = i \bar \Psi^a \partial \llap / \Psi_a  - \frac{1}{\sqrt{2}}
  M_0 \sqrt{G} H \bar t^a t_a - \frac{1}{2} M^2_0 H^2 \ ,
\label{TSM1}
\end{equation}
where $H$ is the real lower component of an auxiliary complex isodoublet
field $\Phi$, and $M_0$ is the bare mass of the field
$H$ (at $\Lambda$). This field becomes a physical Higgs through quantum
effects. We consider this as a prototype model and do not include electroweak
gauge bosons. Furthermore, we do not include in (\ref{TSM1}) the remaining
degrees of freedom of the Higgs isodoublet $\Phi$ -- they
represent (after the condensation) the non-physical Goldstones which are
gauged away into the longitudinal parts of the gauge bosons in
the unitary gauge\footnote{
Stated otherwise, we will be looking only at the effects of the purely
scalar part $((G/4) \bar t^a t_a \bar t^b t_b)$ of the 4-fermion interaction
(\ref{TSM}). Only this part is responsible, within the present framework
where the effects of the loops of (massive) electroweak gauge bosons are
neglected, for the condensation of a $\bar t t$-pair into a scalar Higgs H.}.
In a more realistic framework, we should include either Goldstone bosons
or the entire massive electroweak vector bosons. This enlarged problem
requires new computational methods and has not been solved yet. We are
presently studying the next-to-leading order corrections from Goldstone
bosons and hope to present their impact in the future. We have also studied
modifications from QCD, which are described in Section 3.

We simplify the notation by defining
\begin{equation}
{\sigma} = \frac{M_0}{\sqrt{2}} H \ , \quad {\lambda} = \sqrt{G} \quad
\Longrightarrow \quad
{\cal{L}} = i \bar \Psi^a \partial \llap / \Psi_a - {\lambda}{\sigma}
 \bar t^a t_a - {\sigma}^2 \ .
\label{def1}
\end{equation}
The effective potential is the energy density of the ground state when the
order parameter $\sigma_0 = \langle \sigma \rangle$ is kept fixed.
In the path integral formulation, this
condition is incorporated by inserting a $\delta$-function in the generating
functional for the Lagrangian~\cite{higa},~\cite{kyria}
\begin{equation}
\exp \left[-\frac{i}{\hbar}\Omega V_{\mbox{\footnotesize eff}}({\sigma}_0)
\right] = Z_0^{-1}
\int {\cal{D}} {\sigma} \int {\cal{D}} \bar \Psi {\cal{D}} \Psi
\delta \left[ \int \left( {\sigma}(x)-{\sigma}_0 \right) d^4x \right]
\exp \left[ \frac{i}{\hbar} \int {\cal{L}} d^4x \right] \ ,
\label{pathint1}
\end{equation}
where $\Omega$ is the 4-dimensional volume (formally infinite) and
${\sigma}_0$ is a constant value of the ${\sigma}(x)$-field.
In the generating functional $Z_0$, mentioned above,
\begin{equation}
Z_0 = \int {\cal{D}} {\sigma} \int {\cal{D}} \bar \Psi {\cal{D}} \Psi
 \exp \left[ \frac{i}{\hbar} \int {\cal{L}} d^4x \right] \
\label{Z0}
\end{equation}
appears the Lagrangian of eq.~(\ref{TSM1}), and is independent of
$\sigma_0$. The $\delta$-function can be written in its exponential form
\begin{equation}
\delta(\theta) = (2 \pi)^{-1} \int_{-\infty}^{+\infty} dJ e^{-i \theta J}
\ , \qquad \mbox{where: } \ \theta = \int d^4x ({\sigma}(x)-{\sigma}_0) \ ,
\label{delta}
\end{equation}
which brings the fields $\sigma$ and $\sigma_0$ to the exponential. Then,
going to the Euclidean space by performing Wick's rotation
($i d^4x \mapsto d^4 \bar x$) and using the units in which $\hbar=1$,
we rewrite (\ref{pathint1})
\begin{eqnarray}
\lefteqn{ \exp[-\Omega V_{\mbox{\footnotesize eff}}({\sigma}_0)] =
\mbox{ const} \times \int_{-\infty}^{+\infty} dJ \int {\cal{D}} {\sigma}
 \int {\cal{D}} \bar \Psi {\cal{D}} \Psi \times }
 \nonumber\\
 & & \times \exp \left\{  - \int d^4 \bar x \left[
 -i \bar \Psi^a \bar{\partial \llap /} \Psi_a  +
{\lambda}{\sigma}(\bar x) \bar t^a t_a + {\sigma}(\bar x)^2 +
 i J \left( {\sigma}(\bar x)-{\sigma}_0 \right) \right] \right\} \ .
\label{pathint2}
\end{eqnarray}
The bars over space-time components and over derivatives indicate the
Euclidean quantities~\footnote{
Factor $i$ survives at $J$ in the exponent in (\ref{pathint2}). This is so
because the $\delta$-function constraint, which excludes the contributions
of the constant modes of $\sigma - \sigma_0$, must remain valid also in the
Euclidean metric.}
, $\Omega$ is the Euclidean 4-dimensional volume,
``const'' is a ${\sigma}_0$-independent quantity. Note that
$\bar{\partial \llap /} =  {\partial \llap /}$ (${\bar{\gamma}}^0
= i \gamma^0, {\bar{\gamma}}^j = \gamma^j$). The form (\ref{pathint2})
is convenient for Gaussian integrations. We integrate out
the fermionic degrees of freedom
\begin{equation}
\int {\cal{D}} \bar \Psi  {\cal{D}} \Psi  \exp \left\{ - \int d^4
\bar x \left[ -i \bar \Psi^a \bar{\partial \llap /} \Psi_a  +
{\lambda}{\sigma} \bar t^a t_a \right] \right\}
= \mbox{ det} \hat B[{\sigma}] =
\exp \left\{ Tr \ln \hat B [{\sigma}] \right\} \ ,
\label{pathint3}
\end{equation}
where the operator $\hat B$ is defined
by its matrix element in the $\bar x$-basis
\begin{equation}
\left\langle \bar{x^{\prime}}; j,a | \hat B[{\sigma}] | \bar x; k,b
\right\rangle =
\delta_{j k} \delta_{a b} \left[ i {\bar \gamma}^{\mu}
\frac{\partial}{\partial {\bar x}^{\mu}} + {\lambda}{\sigma}(\bar x)
\delta_{j1} \right] \delta^{(4)}( \bar x - \bar{x^{\prime}} ) \ ,
\label{hatB}
\end{equation}
the indices $j$, $k$ ($=1,2$) being the isospin and $a$, $b$ ($=1,2,3$)
the color indices. We are searching for the effective potential
at values $\sigma_0$ of the scalar field close to the minimum.
Fluctuations around this clasical value $\sigma_0$ are denoted by
$\sigma_1(\bar x)$, which, of course, must be small
\begin{equation}
{\sigma}(\bar x) = {\sigma}_0 + {\sigma}_1(\bar x) \ ,
\qquad \left( {\sigma}_0 = \frac{M_0}{\sqrt{2}} H_0 \right) \ .
\label{omega}
\end{equation}
We can now perform in (\ref{pathint3}) the expansion
in powers of ${\sigma}_1(\bar x)$
\begin{eqnarray}
\ln \hat B[{\sigma}] & = & \ln \hat B [{\sigma}_0] + \ln \left[
1 + {\lambda} \delta_{j1} \hat B [{\sigma}_0]^{-1} {\hat {\sigma}_1} \right]
\nonumber\\
& = & \ln \hat B [{\sigma}_0] + \delta_{j1} \left\{ {\lambda}
\hat B [{\sigma}_0]^{-1} {\hat {\sigma}_1} - \frac{1}{2} {\lambda}^2
\left( \hat B [{\sigma}_0]^{-1} {\hat {\sigma}_1} \right)^2 + \cdots \right\}
\ .
\label{lnB}
\end{eqnarray}
Here, the dots represent quantum fluctuations of higher order than quadratic.
The ${\hat {\sigma}_1}$-operator is defined through: $\langle \bar{x^{\prime}}
| {\hat {\sigma}_1} | \bar x \rangle =$ ${\sigma}_1(\bar x) \delta^{(4)}
(\bar x - \bar{x^{\prime}})$.
The terms linear in ${\sigma}_1$ do not contribute to
$V_{\mbox{\footnotesize eff}}({\sigma}_0)$, due to the $\delta$-function in
(\ref{pathint1}).
Hence, eqs.~(\ref{pathint2})-(\ref{lnB}) yield
\begin{eqnarray}
\lefteqn{ \exp [ - \Omega V_{\mbox{\footnotesize eff}}({\sigma}_0) ]  =
\mbox{ const} \times
\exp \left\{ - \int d^4 \bar x \lbrack {\sigma}_0^2 \rbrack +
Tr \ln \hat B[{\sigma}_0] \right\}
\times }
\nonumber\\
& & \times \int_{-\infty}^{+\infty} dJ \int {\cal{D}} {\sigma}_1
\exp \left\{ - \int d^4 \bar x {\sigma}_1(\bar x)^2
- \frac{1}{2} {\lambda}^2 Tr \left[ ( \hat B [{\sigma}_0]^{-1}
{\hat {\sigma}_1})^2 \right]
- J \int d^4 \bar x {\sigma}_1(\bar x) \right\} \ ,
\label{pathint4}
\end{eqnarray}
where we neglected the effects of cubic and higher quantum fluctuations.
The first factor in (\ref{pathint4}) gives us the usual tree level
and 1-loop contribution to $V_{\mbox{\footnotesize eff}}$
\begin{equation}
(V^{(0)}_{\mbox{\footnotesize eff}}  +
V^{(1)}_{\mbox{\footnotesize eff}} ) ({\sigma}_0) = {\sigma}_0^2
 - \frac{1}{\Omega} Tr \ln \hat B[{\sigma}_0]
 =  {\sigma}_0^2 - \frac{N_{\mbox{\footnotesize c}}}{(2\pi)^4}
 tr_{\mbox{\scriptsize f}}
\int d^4 \bar k \ln (\bar{\gamma^{\mu}} \bar{k^{\mu}}
 + {\lambda} {\sigma}_0 ) \ .
\label{1loop1}
\end{equation}
The integral on the r.h.s.~is derived in Appendix A.
The factor $N_{\mbox{\footnotesize c}}$ comes from the trace over colors,
and the remaining $tr_{\mbox{\scriptsize f}}$ denotes
tracing in the 4-dimensional spinor space.
By expanding the logarithm in (\ref{1loop1}) in
powers of $[{\lambda} {\sigma}_0 (\bar{\gamma^{\mu}} \bar{k^{\mu}})^{-1}]$,
we obtain
\begin{equation}
(V^{(0)}_{\mbox{\footnotesize eff}}  +
V^{(1)}_{\mbox{\footnotesize eff}} ) ({\sigma}_0)
 =  {\sigma}_0^2
 - \frac{N_{\mbox{\footnotesize c}}}{(8 \pi^2)} \int_{0}^{\infty}
 d \bar k^2 \bar k^2
  \ln \left[ 1 + \frac{ {\lambda}^2 {\sigma}_0^2 }{ \bar k^2} \right] \ .
\label{1loop2}
\end{equation}
The expression (\ref{1loop2}) can also be obtained by calculating the
relevant 1-PI diagrams with one loop of the top quark and
zero momentum Higgs particles as the outside legs. This calculation
is described, for example, in ref.~\cite{gceap} where the notation
of eq.~(\ref{TSM1}) was used.

To go beyond one loop, we must integrate the path integral in
(\ref{pathint4}) containing the quantum fluctuations ${\sigma}_1^2$ and
$J {\sigma}_1$ of the scalar in the exponent.
This can be done, by using known formulas for path integrals of
Gaussian distributions (e.g.~ref.~\cite{bailin}). We consider the
general expression
\begin{eqnarray}
\lefteqn{ \int {\cal{D}} {\sigma}_1 \exp \left\{ - \frac{1}{2}
 \int d^4 \bar x d^4 \bar{x^{\prime}}
 {\sigma}_1 (\bar{x^{\prime}}) \hat A(\bar{x^{\prime}}, \bar x)
{\sigma}_1(\bar x) - i J \int d^4 \bar x {\sigma}_1(\bar x) \right\} = }
 \nonumber\\
& & \exp \left[ - \frac{1}{2} Tr \ln \hat A \right]
 \exp \left[ - \frac{1}{2} J^2 \int d^4 \bar x d^4 \bar{x^{\prime}}
\hat A^{-1}(\bar{x^{\prime}}, \bar x) \right] \ ,
\label{pathomega}
\end{eqnarray}
where~\footnote{
Formula (\ref{pathomega}) can be obtained from its well-known
special case of $J=0$ by a simple substitution of variables:
${\hat \sigma}_1 \mapsto {\hat \sigma}_1 - i J {\hat A}^{-1}$.
Furthermore, an identity has to be used whose 1-dimensional
analogue reads: $\int_{-\infty}^{+\infty} dz \exp[-z^2] =
\int_{-\infty}^{+\infty} exp[-(z+iy)^2]$, where $y$ is any real constant.
This identity applies also in our case, because ${\hat A}$ and
${\hat A}^{-1}$ are symmetric and real in the ${\bar x}$-basis, as
shown in Appendix A and this Section.}
for our case
\begin{equation}
{\hat A}(\bar{x^{\prime}},{\bar x}) = 2\delta(\bar{x^{\prime}}- {\bar x} )
 + \lambda^2 tr {\Big \{}
 \langle \bar x|{\hat B}[\sigma_0]^{-1} | \bar{x^{\prime}} \rangle
 \langle \bar{x^{\prime}}|{\hat B}[\sigma_0]^{-1} | \bar x \rangle
{\Big \}} \ ,
\label{Axbas}
\end{equation}
as inferred from (\ref{pathint4}). Here, $tr$ stands for the trace over
the color, isospin and spinor degrees of freedom.

Furthermore, the integration over $J$ in (\ref{pathint4}) of the exponential
factor containing $J^2$ of (\ref{pathomega}) (i.e., the effect of the
$\delta$-function) can also be explicitly performed
\begin{equation}
\int_{-\infty}^{+\infty} dJ \exp \left[ - \frac{1}{2} J^2
 \int d^4 \bar x d^4 \bar{x^{\prime}} \hat A^{-1}(\bar{x^{\prime}}, \bar x)
  \right] = \sqrt{\frac{2\pi}{\alpha}} \ ,
\label{Jint1}
\end{equation}
\begin{equation}
\mbox{ where} \qquad
\alpha  =  \int d^4 \bar x d^4 \bar{x^{\prime}}
 \hat A^{-1}(\bar{x^{\prime}}, \bar x) =
\frac{\Omega}{\tilde A (\bar p = 0 )}  \ .
\label{Jint2}
\end{equation}
We note that $\alpha > 0$ for all interesting regions of values of
${\lambda} {\sigma}_0$ (see Appendix B), hence the integral in (\ref{Jint1})
is convergent. We obtain from (\ref{pathint4}),
(\ref{pathomega})-(\ref{Jint2}) for the contribution of
quadratic fluctuations to the effective potential
the following expression
\begin{equation}
V^{\mbox{\scriptsize (ntl)}}_{\mbox{\footnotesize eff}} ({\sigma}_0)
= \frac{1}{2\Omega} Tr \ln \hat A
- \frac{1}{2 \Omega} \ln \left[ {\tilde A}\left( {\bar p}=0,
\lambda^2 \sigma_0^2 \right) \right] + \cdots \ ,
\label{res1}
\end{equation}
where the superscript ``ntl'' stands for the ``next-to-leading''
order (beyond one loop) and
the dots represent irrelevant ${\sigma}_0$-independent terms.
The first term on the r.h.s.~of
eq.~(\ref{res1}) is $\Omega$-independent in the limit $\Omega$
($=\int d^4 \bar x$) $\to \infty$ (see below).
$\tilde A (\bar p)$ in the second term will be calculated below, with
a finite energy cut-off $\Lambda$, and this integral is finite. Thus the
second term goes to zero as $\Omega \to \infty$. Hence, this term,
which originates from the $\delta$-function of the path integral
(\ref{pathint1}), drops out. The $\delta$-function in
the path integral (\ref{pathint1}) turns out to have the {\it sole}
effect of ensuring that the quantum fluctuations linear
in ${\sigma}_1(\bar x)$ do not contribute to the effective potential.

The tracing in $Tr \ln {\hat A}$ can be performed in the momentum
basis. For this we need the Fourier transform $\tilde A (\bar p)$
of the operator $\hat A$ (see Appendix A for details)
\begin{displaymath}
\langle \bar{x^{\prime}} | \hat A | \bar x \rangle  =
\hat A(\bar{x^{\prime}}, \bar x) = (2 \pi)^{-4} \int d^4 \bar p
e^{i \bar p (\bar x - \bar{x^{\prime}})} \tilde A (\bar p) \ ,
\qquad \tilde A (\bar p) = 2 [1-2 {\lambda}^2 N_{\mbox{\footnotesize c}}
{\cal{K}}_{H} (\bar p^2, \lambda^2{\sigma}_0^2) ] \ ,
\end{displaymath}
\begin{equation}
{\cal{K}}_{H}  (\bar p^2, \lambda^2 {\sigma}_0^2)   =
 \frac{1}{4} \int_{ {\bar k}^2 \leq \Lambda^2_{\mbox{\scriptsize f}} }
\frac{d^4 \bar k}{(2 \pi)^4} tr_{\mbox{\footnotesize f}}  \left[
\frac{i}{({{\bar k} \llap /}- {\lambda} {\sigma}_0)}
\frac{i}{({{\bar p} \llap /} + {{\bar k} \llap /}
 - {\lambda} {\sigma}_0)} \right] \ .
\label{Aomega}
\end{equation}
The integral over $\bar k$ would be in general over an infinite volume.
However, at the $t \bar t$-condensation scale $E \sim \Lambda$
we expect that a new dynamics cuts off the integral.
We stress that we are working all the time in an effective theory
of eq.~(\ref{TSM}), i.e., in the TSM. For this reason, we introduced
in the above integral an energy cut-off $\Lambda_{\mbox{\footnotesize f}}$
for the fermionic (top quark) Euclidean mometum $\bar k$. For simplicity,
the cut-off was chosen to be spherical. This $\Lambda_{\mbox
{\footnotesize f}}$ is to be recognized as being approximately the
energy $\Lambda$ of (\ref{TSM}) at which the $\bar t t$-condensation
takes place. Similarly, the tracing in $Tr \ln \hat A$ in the momentum
basis involves a
second integral -- over the bosonic (Higgs) momenta $\bar p$.
By the arguments just discussed, we introduce for these momenta
a second spherical cut-off $\Lambda_{\mbox{\footnotesize b}}$,
where $\Lambda_{\mbox{\footnotesize b}} \sim \Lambda$.
The tracing in $Tr \ln \hat A$ in the momentum basis is performed
in Appendix A, by using the expression (\ref{Aomega}).
Then we rescale all the Euclidean 4-momenta
$\{ {\bar k}^2,{\bar p}^2 \} \mapsto \Lambda^2_{\mbox{\footnotesize f}}
\{ {\bar k}^2,{\bar p}^2 \}$ and end up with the following result
\begin{equation}
V^{\mbox{\scriptsize (ntl)}} _{\mbox{\footnotesize eff}} ({\sigma}_0)
= \frac{\Lambda^4_{\mbox{\footnotesize f}} }{2(4\pi)^2}
\int_0^{ \Lambda^2_{\mbox{\scriptsize b}}/\Lambda^2_{\mbox{\scriptsize f}} }
 d\bar p^2 \bar p^2
 \ln \left[ 1 -  a {\cal{J}}_{H}  (\bar p^2, \varepsilon^2) \right] \ ,
\label{res2}
\end{equation}
\begin{displaymath}
\mbox{where we denote:} \qquad
\varepsilon^2 = \frac{ {\lambda}^2 }{ \Lambda^2_{\mbox{\footnotesize f}} }
{\sigma}_0^2  = \frac{ G M_0^2 }{ 2 \Lambda^2_{\mbox{\footnotesize f}} } H_0^2
\ ,  \qquad a = \frac{(G N_{\mbox{\footnotesize c}}
\Lambda^2_{\mbox{\footnotesize f}} )}{(8 \pi^2)} \ ,
\end{displaymath}
\begin{equation}
\mbox{and: } \qquad {\cal{J}}_{H}
(\bar p^2, \varepsilon^2) =
  \frac{16 \pi^2}{\Lambda^2_{\mbox{\footnotesize f}} } {\cal{K}}_{H}
 (\Lambda^2_{\mbox{\footnotesize f}} {\bar p}^2, \lambda^2 {\sigma}_0^2)
 = (\pi)^{-2} \int_{\bar k^2 \leq 1} d^4 \bar k
\frac{[\bar k (\bar p + \bar k) - \varepsilon^2]}{(\bar k^2
 + \varepsilon^2)
[(\bar p + \bar k )^2 + \varepsilon^2 ]} \ .
\label{Jdef}
\end{equation}
This is the main result of the analysis so far, giving a closed
form of the next-to-leading effective potential.
As just argued, we expect: $\Lambda_{\mbox{\footnotesize f}} \sim
\Lambda_{\mbox{\footnotesize b}} \sim \Lambda$, where $\Lambda$ is
from (\ref{TSM}). Our numerical results depend on the ratio
$\Lambda_{\mbox{\footnotesize b}}/\Lambda_{\mbox{\footnotesize f}}$.
However, since we are interested
primarily in the qualitative features of the next-to-leading order
corrections, we will assume the equality of the two cut-offs
\begin{equation}
\Lambda_{\mbox{\footnotesize f}} =
\Lambda_{\mbox{\footnotesize b}} = \Lambda \ .
\label{cutoffs}
\end{equation}
The parameter $a$ in (\ref{res2})
can be easily expressed through the ratio $z_1 = (m_{t} ^{(1)}/\Lambda)^2$,
$m_{t} ^{(1)}$ being the solution of the 1-loop gap equation for the mass
of the top quark. The 1-loop gap equation follows from
\begin{displaymath}
\frac{\partial}{\partial \varepsilon^2}
\left( V^{(0)}_{\mbox{\footnotesize eff}}
+ V^{(1)}_{\mbox{\footnotesize eff}}  \right)
{\Big |}_{\varepsilon^2 = z_1}  =  0 \ , \qquad \mbox{ which gives}
\end{displaymath}
\begin{equation}
 a =
 \frac{G N_{\mbox{\footnotesize c}} \Lambda^2}{(8 \pi^2)}
 =  [1 - z_1 \ln(z_1^{-1}+1)]^{-1}
  \ (= {\cal{O}}(1)) \ , \qquad
  \mbox{where: } z_1 = \left( \frac{m_{t} ^{(1)}}{\Lambda} \right)^2  \ .
\label{1lgap}
\end{equation}
The newly introduced dimensionless parameter $a$ is ${\cal{O}}(1)$ ($a>1$).
Furthermore, the integral ${\cal{J}}_{H}  (\bar p^2, \varepsilon^2)$ can be
integrated analytically (see Appendix B). The effective potential has the
following terms
\begin{displaymath}
V_{\mbox{\footnotesize eff}}({\varepsilon}^2) =
 (V_{\mbox{\footnotesize eff}}^{(0)} + V_{\mbox{\footnotesize eff}}^{(1)}
 + V_{\mbox{\footnotesize eff}}^{\mbox{\scriptsize (ntl)}} )
 ( {\varepsilon}^2 )
\end{displaymath}
\begin{equation}
 = \left( \frac{N_{\mbox{\footnotesize c}} \Lambda^4}{8 \pi^2 }
 \right) \left\{ \frac{{\varepsilon}^2}{a}
 - \int_0^1 d {\bar p}^2 {\bar p}^2 \ln (1
 + \frac{{\varepsilon}^2}{{\bar p}^2})
 + \frac{1}{4 N_{\mbox{\footnotesize c}}} \int_0^1 d {\bar p}^2 {\bar p}^2
 \ln \left[ 1 - a {\cal{J}}_{H} ({\bar p}^2, {\varepsilon}^2)
 \right] \right\} \ .
 \label{res3}
 \end{equation}
It is now evident that
$V_{\mbox{\footnotesize eff}}^{\mbox{\scriptsize (ntl)}}$ is
next-to-leading order in the $1/N_{\mbox{\footnotesize c}}$-expansion.

The effective potential can also be derived by the diagrammatic method.
The derivation of $V_{\mbox{\footnotesize eff}}^{(1)}$ is described,
for example, in ref.~\cite{gceap}. The new term
$V_{\mbox{\footnotesize eff}}^{\mbox{\scriptsize (ntl)}}$ is
derived in Appendix C. The relevant diagrams turn out to be those
shown in Fig.~3. It is also shown in the Appendix that the diagrams
of Fig.~3 are all of ${\cal {O}}(1)$ in the
$1/N_{\mbox{\footnotesize c}}$-expansion, and that there are no
other contributing diagrams of ${\cal {O}}(1)$.
An important feature
is the fact that each top quark loop involves an arbitrary number of
external scalar lines with zero momenta.
The $(a {\cal {J}}_H)^{\ell}$-term in the expansion of the last
logarithm in (\ref{res3}) is in fact the contribution of
the $(\ell+1)$-loop diagram of Fig.~3. For example, the first element
in this expansion, $(a {\cal {J}}_H)^1$, is the self-energy
of the scalar particle from the top quark loop, joined together
in a closed ring -- the 2-loop version of Fig.~3. From this
Figure, it becomes evident why we have identified in (\ref{Aomega})
the Euclidean 4-momentum $\bar k$ as the fermionic (top quark)
momentum with a cut-off $\Lambda_{\mbox{\footnotesize f}}$,
and $\bar p$ as the bosonic (Higgs) momentum with a cut-off
$\Lambda_{\mbox{\footnotesize b}}$.

Finally, we emphasize that the diagrammatic method involves
tedious combinatorics and summations. However, its pictorial
transparency makes it possible to adapt it easily to other
interactions, like the QCD interaction between the gluon and
the top quark. The latter results in contributions of QCD
to the effective potential, which are studied in the nect Section.

\section{QCD contributions to the effective potential}

As mentioned already, the dominant part of the gluonic contributions
to $V_{\mbox{\footnotesize eff}}$ can be calculated in a straightforward
way by a generalization of the diagrammatic method,
in complete analogy to the calculation presented in Appendix C for
the contribution of the Higgs sector to $V_{\mbox{\footnotesize eff}}$.
The relevant graphs are those of Fig.~3, where now all the internal dotted
lines represent a gluon of a specific color.
The results (\ref{Vl2})-(\ref{resC}) can basically be transcribed,
if we simply replace the ``Yukawa'' coupling $\sqrt{G}M_0/\sqrt{2}$
$\mapsto$ $g_{\mbox{\scriptsize s}} \lambda^a \gamma_{\mu}/2$
(where: $g_{\mbox{\scriptsize s}}^2/(4\pi) = \alpha_{\mbox
{\footnotesize s}}$; $\lambda^a/2$ are the generators of the
$SU(3)_c$-group). In addition, we replace the non-dynamical
Higgs propagator $(-i)/M_0^2$ by the gluon propagator
$(-i)\delta^{ab} \left( g^{\mu\nu}-p^{\mu}p^{\nu}/p^2 \right)/p^2$
($a$, $b$ are color indices).
For the gluons we use the Landau gauge,
in which there is no (1-loop) QCD contribution to the
quark wave function renormalization.
In close analogy to Appendix C, we obtain for the contribution of the
gluonic diagrams (in the Eucl.~metric)
\begin{equation}
V_{\mbox{\footnotesize eff}}^{\mbox{\scriptsize (gl)}} \left( \sigma_0 \right)
 = \frac{1}{2(4\pi)^2}(N_{\mbox{\footnotesize c}}^2-1)
 \int_0^{\Lambda_{\mbox{\scriptsize b}} ^2}
d{\bar p}^2 {\bar p}^2 \ln \left[ 1 - 2 g_{\mbox{\scriptsize s}}^2
{\cal {K}}_{\mbox{\footnotesize gl}}  \left(
 {\bar p}^2, \lambda^2 \sigma_0^2 \right) \right] \ ,
\label{QCD1}
\end{equation}
\begin{eqnarray}
\mbox{where: } \quad
{\cal {K}}_{\mbox{\footnotesize gl}}  \left( {\bar p}^2,
\lambda^2 \sigma_0^2 \right) & = &
{\cal {K}}^{\mu\nu} \left(
{\bar p}^2,\lambda^2 \sigma_0^2 \right) \frac{1}{{\bar p}^2}
 \left( \delta^{\mu\nu} - \frac{{\bar p}^{\mu}{\bar p}^{\nu}}{{\bar p}^2}
  \right) \ ,
 \nonumber\\
{\cal {K}}^{\mu\nu} \left( {\bar p}^2,\lambda^2 \sigma_0^2 \right) & = &
 \frac{1}{4} \int \frac{d^4{\bar k}}{(2\pi)^4}
 tr_{\mbox{\footnotesize f}}  \left[
 {\bar \gamma}^{\mu} \frac{i}{
 \left( {\bar {k \llap /}}- \lambda\sigma_0 \right)}
  {\bar \gamma}^{\nu} \frac{i}{
  \left( {\bar {p \llap /}} + {\bar {k \llap /}} - \lambda\sigma_0 \right) }
     \right] \ .
 \label{QCD2}
 \end{eqnarray}

 We point out that the contributions of those diagrams of Fig.~3
 which contain simultaneously scalar and gluonic internal propagators are
 zero. This is so because in such diagrams there is at least one top quark
 loop which has one scalar and one gluon propagator attached to it, thus
 resulting in a factor proportional to the trace over colors $tr(\lambda^a)
 =0$.

 The integral over the top quark momentum in ${\cal {K}}^{\mu\nu}$
 can be regularized by several methods, provided they respect the
 symmetries of the QCD gauge theory (Lorentz invariance, etc.).
 QCD being a renormalisable theory, the self-energy of the gluons,
 and thus ${\cal {K}}^{\mu\nu}$, have only logarithmic cut-off
 dependence. This is in strong contrast with the scalar sector, which has
 $\Lambda^2$-terms under any regularization - cf.~eq.~(\ref{Aomega}).
 The integral for ${\cal {K}}^{\mu\nu}$ thus obtained is well-known
 in the literature. Using the well-known technique of exponentiating
 the denominators as integrals of exponentials, it
 can be written in the ``proper time'' form (and in Euclidean metric)
 \begin{equation}
 {\cal {K}}^{\mu\nu} \left( {\bar p}^2, \lambda^2\sigma_0^2 \right)
 = \frac{\left( -\delta^{\mu\nu} {\bar p}^2 + {\bar p}^{\mu}{\bar p}^{\nu}
 \right) }{8 \pi^2} \int_0^1 dz z(1-z) \int_0^{\infty} \frac{d\tau}{\tau}
 \rho_{\mbox{\footnotesize f}} (\tau;\Lambda_{\mbox{\footnotesize f}} )
 \exp {\big \{ } -\tau \left[z(1-z){\bar p}^2
 +\lambda^2\sigma_0^2 \right] {\big \} } \ ,
 \label{QCD3}
 \end{equation}
 where $\rho_{\mbox{\footnotesize f}} (\tau;\Lambda_{\mbox
 {\footnotesize f}} )$ is a regulator for the fermionic (top
 quark) momenta, with some effective cut-off
 $\Lambda_{\mbox{\footnotesize f}} $. Note that $\rho_{\mbox{\footnotesize f}}
 =1$ for $\tau \gg 1/\Lambda_{\mbox{\footnotesize}}$, and
 $\rho_{\mbox{\footnotesize f}}=0$ for  $\tau \ll
 1/\Lambda_{\mbox{\footnotesize}}$.
 The regulator eliminates the $\Lambda^2_{\mbox{\footnotesize f}}$-term
 in ${\cal {K}}^{\mu\nu}$. Among the many choices,
 we select two different regulators $\rho_{\mbox{\footnotesize f}} $:
 the simple Pauli-Villars (P.V.) subtraction;
 and the ``proper time'' cut-off ($\rho_{\mbox{\footnotesize f}}
 (\tau;\Lambda_{\mbox{\footnotesize f}} ) =$
 $\theta(\tau - 1/\Lambda_{\mbox{\footnotesize f}} ^2)$)
 \begin{equation}
 {\cal {K}}^{\mu\nu}_{\mbox{\scriptsize P.V.}}
 \left({\bar p}^2, \lambda^2\sigma_0^2 \right)
 ={\cal {K}}^{\mu\nu}\left({\bar p}^2, \lambda^2\sigma_0^2 \right) -
  {\cal {K}}^{\mu\nu}\left({\bar p}^2,
  \Lambda_{\mbox{\footnotesize f}} ^2 \right) \ ,
  \label{QCDPV}
  \end{equation}
  \begin{equation}
 {\cal {K}}^{\mu\nu}_{\mbox{\footnotesize p.t.}}\left({\bar p}^2,
 \lambda^2\sigma_0^2 \right) =
 \frac{1}{8\pi^2} \left(-\delta^{\mu\nu}{\bar p}^2
 +{\bar p}^{\mu}{\bar p}^{\nu} \right) \int_0^1 dz z (1-z)
 \int_{1/\Lambda_{\mbox{\footnotesize f}} ^2}^{\infty}
 \frac{d\tau}{\tau} \exp {\big \{ }
 -\tau \left[ z(1-z){\bar p}^2+\lambda^2\sigma_0^2
 \right] {\big \} } \ .
 \label{QCDpt}
 \end{equation}
 Both of these integrals can be integrated analytically. This then leads
 to the following dominant part of the gluonic contributions~\footnote{
  Just like in Section 2, we equate for simplicity the bosonic and the
  fermionic cut-offs: $\Lambda_{\mbox{\footnotesize b}}  =$
  $\Lambda_{\mbox{\footnotesize f}}  =$$ \Lambda$;
  and rescale the momenta and
  the masses (${\bar q}^2 \mapsto$$ {\bar q}^2/\Lambda^2$;
  ${\varepsilon}^2 =$$ \lambda^2 \sigma_0^2/\Lambda^2$). }
  to $V_{\mbox{\footnotesize eff}}$
 \begin{equation}
 V_{\mbox{\footnotesize eff}}^{\mbox{\scriptsize (gl)}}
 \left( {\varepsilon}^2 \right)
   = \frac{\Lambda^4}{2 (4\pi)^2}
   \left( N_{\mbox{\footnotesize c}}^2-1 \right)
 \int_0^1 d{\bar p}^2 {\bar p}^2 \ln \left[ 1 - a_{\mbox{\footnotesize gl}}
  {\cal {J}}_{\mbox{\footnotesize gl}}
 \left( {\bar p}^2,{\varepsilon}^2 \right) \right]  \ ,
 \label{QCDres}
 \end{equation}
 \begin{eqnarray}
 \mbox{where: } \quad
 {\cal {J}}_{\mbox{\footnotesize gl}}
 \left( {\bar p}^2,{\varepsilon}^2 \right) & = &
 \frac{8 \pi^2}{3} {\cal {K}}_{\mbox{\footnotesize gl}}
 \left( \Lambda^2 {\bar p}^2, \lambda^2\sigma_0^2 =
 \Lambda^2 {\varepsilon}^2 \right) \ ,
 \nonumber\\
 a_{\mbox{\footnotesize gl}}  & = & \frac{3}{4 \pi^2}g_{\mbox{
 \scriptsize s}}^2
 = \frac{3 \alpha_{\mbox{\footnotesize s}}}{\pi} \approx
 0.105 \qquad (\mbox{for } \alpha_{\mbox{\footnotesize s}} =
 \alpha_{\mbox{\footnotesize s}}(m_{t} ) \approx 0.11) \ .
 \label{QCDresdef}
 \end{eqnarray}
 The explicit expressions for $J_{\mbox{\footnotesize gl}} $ for
 the case of one simple
 Pauli-Villars subtraction and for the case of the ``proper time''
 cut-off are given at the end of Appendix B (see eqs.~(\ref{JglPV})
 and (\ref{Jglpt})). The expression (\ref{QCDres}) represents in
 a sense the ``leading-logarithmic'' QCD contribution to
 $V_{\mbox{\footnotesize eff}}$. Evidently, it includes
 the two loop contribution, represented by the following simple diagram
 integrated over the quark and gluon momenta: ${\bar t} t$-loop
 and a gluon propagator across it (2-loop version of Fig.~3).
 This leading two-loop contribution of QCD to $V_{\mbox{\footnotesize eff}}$
 is obtained from (\ref{QCDres}) by replacing
 $\ln \left[1 - a_{\mbox{\footnotesize gl}}
 {\cal {J}}_{\mbox{\footnotesize gl}}
 \left( {\bar p}^2,{\varepsilon}^2 \right) \right] $ by its leading
 term $\left[- a_{\mbox{\footnotesize gl}}
 {\cal {J}}_{\mbox{\footnotesize gl}}
 \left( {\bar p}^2,{\varepsilon}^2 \right) \right]$.

 It is not possible to identify the expression (\ref{QCDres}) as
 corresponding to a next-to-leading contribution in the
 $1/N_{\mbox{\footnotesize c}}$-expansion of $V_{\mbox{\footnotesize eff}}/
 N_{\mbox{\footnotesize c}}$ in the case of the
 {\it formal} limit $N_{\mbox{\footnotesize c}} \to \infty$.
 Unlike the expression (\ref{res2}), or the last term in (\ref{res3}),
 it contains also the factor
 $(N_{\mbox{\footnotesize c}}^2-1)$, i.e., the number of gluons
 (cf.~also eq.~(\ref{res4}) below). It is not clear whether the QCD
 contributions can be organized in a systematic way into an
 $1/N_{\mbox{\footnotesize c}}$-expansion series whose $n$'th expansion
 term would have, for example, a factor $(N^2_{\mbox{\footnotesize c}}-1)
 /N^n_{\mbox{\footnotesize c}}$ in front of it. Such a series would
 probably be for $N_{\mbox{\footnotesize c}}=3$ a convergent or at least
 an asymptotic series, whose first term would be (\ref{QCDres}).
 We point out that the $1/N_{\mbox{\footnotesize c}}$-expansion (with
 $N_{\mbox{\footnotesize c}}=3$) of $V_{\mbox{\footnotesize eff}}$ in the
 present paper refers to the contributions from the electroweak (or: scalar,
 Higgs) sector only, i.e., to the contributions corresponding to the
 diagrams which contain the top quarks and composite scalars, but no gluons.
 $N_{\mbox{\footnotesize c}}$($=3$) represents in this case the
 number of top quark types contributing in loops separately to the
 condensation. The related expansion of these contributions in powers of
 $1/3$ ($=1/N_{\mbox{\footnotesize c}}$), as discussed in the previous
 Section and in Appendix C, should therefore be regarded as independent of
 a particular structure of QCD and of the number of gluons. On the other
 hand, what is the most simple way to organize the QCD contributions to
 $V_{\mbox{\footnotesize eff}}$, so that they would result in a convergent
 or at least an asymptotic series? Such a series appears to be the direct
 loop expansion in powers of $\alpha_s$. In this context, we point out that
 expression (\ref{QCDres}), unlike (\ref{res2}), turns out to be almost
 equal to the two-loop
 approximation, i.e., to the first term from the expansion of the logarithm,
 the difference being only a fraction of a percent. This is connected to the
 fact that, at the considered scales ($m_t<E<\Lambda$), QCD can be treated
 perturbatively. The smallness of the couplings $\alpha_s$ (or:
 $a_{\mbox{\footnotesize gl}}$) at these scales makes the total gluonic
 contribution $V_{\mbox{\footnotesize eff}}^{\mbox{\scriptsize gl}}$
 small despite a relatively large coefficient $N^2_{\mbox{\footnotesize c}}
 -1 = 8$.

\section{Analysis of the improved gap equation}

Summarizing the results of the previous sections, we write down
the effective potential which includes the (0+1)-loop contributions,
the next-to-leading
contributions ($V_{\mbox{\footnotesize eff}}^{\mbox{\scriptsize (ntl)}}$)
and the dominant gluonic contributions
($V_{\mbox{\footnotesize eff}}^{\mbox{\scriptsize gl}} $)
\begin{displaymath}
V_{\mbox{\footnotesize eff}}({\varepsilon}^2) =
 \left( V_{\mbox{\footnotesize eff}}^{(0)}
 +V_{\mbox{\footnotesize eff}}^{(1)}
 + V_{\mbox{\footnotesize eff}}^{\mbox{\scriptsize (ntl)}}
 + V_{\mbox{\footnotesize eff}}^{\mbox{\scriptsize (gl)}}  \right)
 ( {\varepsilon}^2 )
\end{displaymath}
\begin{eqnarray}
& = & \left( \frac{N_{\mbox{\footnotesize c}} \Lambda^4}{8 \pi^2 }
\right) {\Big \{ } \frac{{\varepsilon}^2}{a}
 - \int_0^1 d {\bar p}^2 {\bar p}^2 \ln (1
 + \frac{{\varepsilon}^2}{{\bar p}^2})
 + \frac{1}{4 N_{\mbox{\footnotesize c}}} \int_0^1 d {\bar p}^2 {\bar p}^2
 \ln \left[ 1 - a {\cal{J}}_{H} ({\bar p}^2, {\varepsilon}^2) \right]+
 \nonumber\\
 & & + \frac{(N_{\mbox{\footnotesize c}}^2-1)}{4 N_{\mbox{\footnotesize c}}}
 \int_0^1 d {\bar p}^2 {\bar p}^2
 \ln \left[ 1 - a_{\mbox{\footnotesize gl}}
 {\cal{J}}_{\mbox{\footnotesize gl}} ({\bar p}^2, {\varepsilon}^2) \right]
 {\Big \} } \ .
 \label{res4}
 \end{eqnarray}

We obtain an improved
gap equation by minimizing this entire effective potential.
This gap equation gives us
an improved value $z=(m_{t} /\Lambda)^2$
as a function of the input parameter $(G \Lambda^2)$ ($=8 \pi^2 a/
N_{\mbox{\footnotesize c}}$)
\begin{displaymath}
 {\Big \{ }
\frac{1}{a} - \int_0^1 \frac{ d {\bar p}^2 {\bar p}^2 }{ {\bar p}^2
+{\varepsilon}^2 } + \frac{1}{4 N_{\mbox{\footnotesize c}}}
\frac{\partial}{ \partial {\varepsilon}^2 }
\int_0^1 d {\bar p}^2 {\bar p}^2 \ln \left[ 1 - a {\cal{J}}_{H}
({\bar p}^2, {\varepsilon}^2)
\right] +
\end{displaymath}
\begin{equation}
+ \frac{(N_{\mbox{\footnotesize c}}^2-1)}{4 N_{\mbox{\footnotesize c}}}
\frac{\partial}{ \partial {\varepsilon}^2 }
\int_0^1 d {\bar p}^2 {\bar p}^2 \ln \left[ 1 -
a_{\mbox{\footnotesize gl}}  {\cal{J}}_{\mbox{\footnotesize gl}}
({\bar p}^2, {\varepsilon}^2) \right]
{\Big \} } {\Big |}_{ {\varepsilon}^2=z } = 0  \ .
\label{2lgap}
\end{equation}
In Appendix B we give analytic expressions for ${\cal{J}}_{H}
({\bar p}^2, {\varepsilon}^2)$ and $\partial {\cal{J}}_{H}
(\bar p^2, \varepsilon^2)/\partial \varepsilon^2$ (spherical
cut-off), as well as for ${\cal{J}}_{\mbox{\footnotesize gl}}
({\bar p}^2, {\varepsilon}^2)$ and implicitly for
$\partial {\cal{J}}_{\mbox{\footnotesize gl}}
(\bar p^2, \varepsilon^2)/\partial \varepsilon^2$ (for the case
of the simple Pauli-Villars subtraction, and for the case of
the ``proper time'' cut-off). Using these expressions in (\ref{2lgap}),
we calculate the integrals numerically (for a
given $a$ and for $N_{\mbox{\footnotesize c}}=3$) and obtain
solutions for ${\varepsilon}^2 = z=
(m_{t} /\Lambda)^2$ as a function of the input parameter $a$,
that is, of the input parameter $G \Lambda^2 = 8 \pi^2
a/N_{\mbox{\footnotesize c}}$.
The results are tabulated in the third column (with
gluonic contributions included) and in the fifth column (without gluonic
contributions) of Table 1, and are depicted in Figs.~1a-1e
(for specific input values $G \Lambda^2$) as full lines and long-dashed
lines, respectively. For gluons
we chose the ``proper time'' cut-off (\ref{QCDpt}). The results for
$\sqrt{z} = m_{t} /\Lambda$ in the third column of Table 1
change mostly by less than 3 percent if we choose
for the gluonic contributions the simple Pauli-Villars subtraction
(\ref{QCDPV}) instead.

When solving eq.~(\ref{2lgap}) for $z=z(G\Lambda^2)$, we must always
check: that the integrands in (\ref{2lgap}) and (\ref{res4})) have
no singularities near the minimum ${\varepsilon}^2 \approx z$
and, in addition, that the stationary point ${\varepsilon}^2=z$ of
$V_{\mbox{\footnotesize eff}}({\varepsilon}^2)$ is really a local
minimum. The calculations show that there are no singularities for
${\varepsilon}^2$ lying near the value of $z$ for the minimum of
$V_{\mbox{\footnotesize eff}}$ (cf.~discussion in Appendix B).
Furthermore, the calculations show an interesting phenomenon:
as $a=(N_{\mbox{\footnotesize c}} G \Lambda^2)/(8 \pi^2)$  decreases
beyond a critical
value of approximately $1.50$ ($1.60$ without gluons), the local
minimum of the effective potential $V_{\mbox{\footnotesize eff}}
(\varepsilon^2)$ disappears
at a {\it non-zero} value $\varepsilon^2 = 0.277^2 \approx 0.08$
($0.309^2 \approx 0.095$ without gluons),
i.e., the dynamical symmetry breaking disappears (see Figs.~1d-1e).
This is also seen in Table 1 (first, third and fifth
column).
Hence, the calculations suggest an upper bound on the energy cut-off
$\Lambda$ ($\simeq E_{\mbox {\footnotesize condensation}}$),
and this bound is expressed in
terms of the bare mass $m_{t} $
\begin{equation}
\frac{m_{t} }{\Lambda} \ \stackrel{>}{\sim} 0.277 \ (0.309) \qquad
\Rightarrow \qquad
 \Lambda \ \stackrel{<}{\sim} \ 3.61 m_{t}  \ (3.24 m_{t} ) \ .
\label{Lbound}
\end{equation}
Here and for the rest of this paper, the values in parentheses indicate
values when the QCD contributions
are ignored. At the same time, these numerical results give a lower
bound on the 4-fermion coupling $G$ of eq.~(\ref{TSM})
\begin{equation}
a=\frac{N_{\mbox{\footnotesize c}} G \Lambda^2}{8 \pi^2} \
\stackrel{>}{\sim} \ a_{\mbox{\footnotesize crit.}} = 1.499 \ (1.597) \ .
\label{Gbound}
\end{equation}
We note that the 1-loop gap equation (\ref{1lgap}) gives a much weaker
bound on $G$, and {\it no} relation analogous to (\ref{Lbound})
\begin{displaymath}
a = \frac{N_{\mbox{\footnotesize c}} G \Lambda^2}{8 \pi^2}
\stackrel{>}{\sim}  1 \ , \qquad
\mbox{and } \quad \Lambda \leq \infty  \ \ (\mbox{or: }
\Lambda \stackrel{<}{\sim} E_{\mbox {\footnotesize Planck}}) \ .
\end{displaymath}

We also note from Table 1 (third column) that,
as $G$ goes down to $G_{\mbox{\footnotesize crit.}} \simeq
1.50 \times 8 \pi^2/(N_{\mbox{\footnotesize c}} \Lambda^2)$
(for $N_{\mbox{\footnotesize c}}=3$),
the difference between $(m_{t} /\Lambda)$ and the 1-loop solution
$(m_{t} ^{(1)}/\Lambda)$ becomes as large as 34 percent.
We can also take the following point of view: if we require that
the calculated corrections beyond 1-loop do not change the
solution $z_1 = (m_{t} ^{(1)}/\Lambda)^2$ of the 1-loop gap equation
drastically~\footnote{
It is manifestly apparent from the expression for
$V_{\mbox{\footnotesize eff}}$ that
the natural parameter that should be regarded as the solution of
the gap equation is $z=(m_{t} /\Lambda)^2$, and not
$\sqrt{z}=(m_{t} /\Lambda)$.},
e.g.~not by more than a factor of $2$, we
arrive at very similar results to those of
eqs.~(\ref{Lbound})-(\ref{Gbound})
\begin{equation}
\frac{z}{z_1} \ > \ 0.5 \ \ \Rightarrow \ \  \frac{m_{t} }{\Lambda} \
\stackrel{>}{\sim} \ 0.299 \ (0.332) \ ,
\ a \ \stackrel{>}{\sim} \ 1.506 \ (1.605) \ .
\label{boundsalt}
\end{equation}

Furthermore, singularities (cuts) do occur in
$V_{\mbox{\footnotesize eff}}^{\mbox{\scriptsize (ntl)}} $
in the region of small ${\varepsilon}^2$ (${\varepsilon}^2 < 0.25 z_1$,
for all $a \geq a_{\mbox{\footnotesize crit.}}$), and the (full)
lines in Figs.~1a-1e were not continued into these regions.
In these regions, the expression in the logarithm of the
integrand for $V_{\mbox{\footnotesize eff}}^{\mbox{\scriptsize (ntl)}} $
becomes negative (cf.~Appendix B for discussion). The loop
expansion of $V_{\mbox{\footnotesize eff}}^{\mbox{\scriptsize (ntl)}} $
(i.e., in powers of
$[a {\cal{J}}_{H} ({\bar p}^2,{\varepsilon}^2)]$, cf.~eqs.~(\ref{Vl3}),
(\ref{resC})) is a divergent series for such small ${\varepsilon}^2$.
These regions of ${\varepsilon}^2$ do not represent any problem,
since the region of our interest is around the
minimum of $V_{\mbox{\footnotesize eff}}$ (${\varepsilon}^2 \approx z$)
which lies a safe distance away from the singularities,
even in the critical case $a \approx 1.50$ ($z \approx 0.277^2$,
cf.~Figs.~1d-1e). Stated otherwise: as long as we require that the
calculated next-to-leading corrections to the 1-loop gap solution
$z_1$ do not change this solution drastically ($e.g., z > 0.3 z_1$),
i.e., that the $1/N_{\mbox{\footnotesize c}}$-expansion of the scalar
contributions (for $N_{\mbox{\footnotesize c}}=3$) make sense,
we do not encounter any singularities near the minimum of
$V_{\mbox{\footnotesize eff}}^{\mbox{\scriptsize (ntl)}} $.

The gluonic contributions $V_{\mbox{\footnotesize eff}}
^{\mbox{\scriptsize (gl)}} $ of eq.~(\ref{QCDres}) do
not have any of these problems, because
${\cal {J}}_{\mbox{\footnotesize gl}} ({\bar p}^2, {\varepsilon}^2)$
turns out to be negative in the entire region of integration.
In addition, this ${\cal {J}}_{\mbox{\footnotesize gl}} $ is
numerically very small, and the
corresponding logarithm in (\ref{QCDres}) is in addition suppressed
by the small coupling $a_{\mbox{\footnotesize gl}} \approx
3 \alpha_{\mbox{\footnotesize s}}(m_{t} )/\pi \approx 0.105$.
The smallness of $a_{\mbox{\footnotesize gl}} $ is connected
with the perturbative nature
of QCD at our relevant energies $E \stackrel{>}{\sim}
m_{t} $. Therefore, the contributions of gluons are playing a
numerically inferior role when compared to those of the next-to-leading
Higgs contributions to the gap equations (cf.~Table 1 and
eqs.~(\ref{Lbound})-(\ref{Gbound})). A consequence of the
relative small gluonic contributions is the fact that
the results of this paper are practically the same
when $V_{\mbox{\footnotesize eff}}^{\mbox{\scriptsize (gl)}} $ in
(\ref{QCDres}) is replaced by the 2-loop gluonic contribution, i.e.,
when we replace $\ln \left[1 - a_{\mbox{\footnotesize gl}}
{\cal {J}}_{\mbox{\footnotesize gl}}
 \left( {\bar p}^2,{\varepsilon}^2 \right) \right] $ by its leading
 term $\left[- a_{\mbox{\footnotesize gl}}
 {\cal {J}}_{\mbox{\footnotesize gl}}
 \left( {\bar p}^2,{\varepsilon}^2 \right) \right]$.
This substitution brings a difference of a fraction of a percent.
 For the next-to-leading Higgs contributions,
 $V_{\mbox{\footnotesize eff}}^{\mbox{\scriptsize (ntl)}} $,
 the situation is different, the 2-loop approximation
 of $V_{\mbox{\footnotesize eff}}^{\mbox{\scriptsize (ntl)}} $
 differs notably from the full version of
 $V_{\mbox{\footnotesize eff}}^{\mbox{\scriptsize (ntl)}} $
(cf.~Figs.~1a-1c, dash-dotted and long-dashed lines).
For illustration, in Figs.~1a-c we included also the effective potential
calculated up to 1-loop (short-dotted line).

\section{Mass renormalization corrections and conclusions}

It should be pointed out that the solution for $m_{t} $ through the
gap equation (\ref{2lgap}) (Table 1, third and fifth column)
is not the {\it physical} mass of the top quark in the present framework,
but rather the ``bare'' mass of the top quark in the effective theory with
an energy cut-off $\Lambda$:
\begin{equation}
m_{t}  = m_{t} ^{\mbox{\footnotesize bare}} = m_{t} (\Lambda)
\label{baremass}
\end{equation}
In other words, we have calculated essentially the contribution of the
tadpoles to $m_{t} $, i.e., the value of the condensate
$\langle \bar \Psi \Psi \rangle$ -- in such a framework, the relation
between the condensate and the physical mass has been investigated by
Politzer~\cite{politzer}.
The physical mass $m_{t} ^{\mbox{\footnotesize ren.}}$ is calculated,
within the present framework, by adding up
the diagrams of Figs.~5a-c. The dashed lines represent the
(composite) Higgs propagator (this results in the next-to-leading
in $1/N_{\mbox{\footnotesize c}}$ correction to $m_{t} $ due to the
scalar sector). In addition, we include the usual 1-loop QCD
correction $(\delta m_{t} )^{\mbox{\scriptsize QCD}}$
which is obtained by the diagram of Fig.~5a, where now the dashed
line represents the gluonic propagator. We do not include the
QCD effects that are higher than this 1-loop QCD renormalization effect.
This is justified, because these higher QCD effects are quite
negligible, as suggested also by the mentioned smallness of the
function $a_{\mbox{\footnotesize gl}}
{\cal {J}}_{\mbox{\footnotesize gl}} ({\bar p}^2,{\varepsilon}^2)$.
The second part of Appendix C
contains details of the calculation of these diagrams. This leads
to the corrected (renormalized) mass of the top quark
(cf.~(\ref{pitrad})-(\ref{dmQCD}))
\begin{eqnarray}
 \frac{m_{t} ^{\mbox{\footnotesize ren.}}}{\Lambda} & = &
 \frac{m_{t} ^{\mbox{\footnotesize bare}}}{\Lambda} +
 \left( \frac{a}{4 N_{\mbox{\footnotesize c}}} \right) \int_0^1
\frac{dw}{\left[ 1 - a {\cal{J}}_{H} (w,x^2) \right]}  {\Big [ }
\frac{1}{4 x^3} w \left( w + 2 x^2 - \sqrt{w(w+4 x^2)} \right) +
\nonumber\\  & &
+ \frac{1}{x} \left( w - \sqrt{w(w+4x^2)} \right) {\Big ] } +
\frac{\left( \delta m_{t}  \right)^{\mbox{\scriptsize QCD}}}{\Lambda} \ ,
\label{mtren}
\end{eqnarray}
where we employed the notation
\begin{displaymath}
x = \frac{m_{t} }{\Lambda} = \frac{m_{t} ^{\mbox{\footnotesize bare}}}
{\Lambda} \ ,
\end{displaymath}
and $w$ is the square of the 4-momentum ${\bar p}/\Lambda$ of the
(non-dynamical) Higgs $H$ appearing in Figs.~5a-c (cf.~(\ref{pitrad})).
The $(\delta m_{t} )^{\mbox{\scriptsize QCD}}$ can be obtained from
(\ref{dmQCD}). For the case of the simple Pauli-Villars subtraction this
yields
\begin{equation}
\left( \delta m_{t}  \right)^{\mbox{\scriptsize QCD}}
_{\mbox{\footnotesize P.V.}} =
\frac{\alpha_{\mbox{\footnotesize s}}}{\pi} m_{t}
\ln \left( \frac{\Lambda^2}{m_{t} ^2} \right) \ ,
\label{dmQCDPV}
\end{equation}
and for the case of the ``proper time'' cut-off this gives
\begin{equation}
\left( \delta m_{t}
\right)^{\mbox{\scriptsize QCD}}_{\mbox{\footnotesize p.t.}} =
\frac{\alpha_{\mbox{\footnotesize s}}}{\pi} m_{t}  \left[
\ln \left( \frac{\Lambda^2}{m_{t} ^2} \right)
+ 0.2561... + \frac{5}{9} \frac{m_{t} ^2}{\Lambda^2} +
{\cal {O}}\left( \frac{m_{t} ^4}{\Lambda^4} \right) \right]   \ .
\label{dmQCDpt}
\end{equation}
The logarithmic terms in (\ref{dmQCDPV}) and (\ref{dmQCDpt}) can be
reproduced by using the 1-loop renormalization group equation of the
Standard Model for the top quark Yukawa coupling, by ignoring all the
terms contributing to the running of that coupling except those of QCD.

The renormalized quantities $(m_{t} ^{\mbox{\footnotesize ren.}}/
\Lambda)$, which correspond to the full bare values $(m_{t} /\Lambda)$
of the third column of Table 1, were calculated from (\ref{mtren})
numerically, using for $(\delta m_{t} )^{\mbox{\scriptsize QCD}}$ again
the ``proper time'' cut-off (\ref{dmQCDpt}).
They are given in the adjacent fourth column of Table 1.
For the case without the QCD contributions (the fifth column),
the corresponding renormalized quantities are in the adjacent sixth column -
calculated from (\ref{mtren}) without the $(\delta m_{t} )^
{\mbox{\scriptsize QCD}}$ term.
The $(0+1)$-loop bare mass $m_{t} ^{(1)}$ (second column)
does not get renormalized at this level, since we remain in this
case only at the leading order level in the
$1/N_{\mbox{\footnotesize c}}$-expansion.

The result (\ref{Lbound}) and Table 1 imply, within our framework,
that the symmetry breaking (condensation) occurs only for
\begin{equation}
\frac{m_{t} ^{\mbox{\footnotesize ren.}}}{\Lambda} \ \stackrel{>}
{\sim} \ 0.217 \ (0.213) \qquad
\Rightarrow \qquad \Lambda \stackrel{<}{\sim} 4.6
m_{t} ^{\mbox{\footnotesize ren.}}
\ (4.7 m_{t} ^{\mbox{\footnotesize ren.}}) \ .
\label{phy1}
\end{equation}
The values in brackets correspond to the case when no QCD effects are
included. For $m_{t} ^{\mbox{\footnotesize ren.}}$ with the value of
$180 {\mbox GeV}$~\cite{CDF}, we
obtain within the present framework the main result of our paper:
there is an upper bound
for the energy cut-off $\Lambda = E_{\mbox{\footnotesize condensation}}$
\begin{equation}
\Lambda \ \stackrel{<}{\sim} \ 0.830 \mbox{ TeV} \ (0.845 \mbox{ TeV}) \ .
\label{phy2}
\end{equation}
The value in the brackets corresponds to the case when the
QCD effects are ignored.
We see that the inclusion of the leading part of the QCD effects
does not affect appreciably this upper bound. Furthermore, if we
use in the QCD contributions instead of the ``proper time'' cut-off
the simple Pauli-Villars subtraction, the results
($a_{\mbox{\footnotesize crit.}} \approx 1.48$,
$m_{t} ^{\mbox{\footnotesize ren.}}/\Lambda \ \stackrel{>}{\sim}
\ 0.215$) differ very little from the ``proper time case'' for the
gluonic contributions ($a_{\mbox{\footnotesize crit.}} \approx 1.50$,
$m_{t} ^{\mbox{\footnotesize ren.}}/\Lambda \
\stackrel{>}{\sim} \ 0.217$).

If we take instead of the requirement of the disappearance of
the minimum (cf.~(\ref{Lbound}))
a more conservative point of view, namely, the
requirement that the calculated corrections to the 1-loop gap
equation solution $z_1$ should not drastically change it
(not by more than factor $2$, cf.~(\ref{boundsalt})), then the resulting
upper bound on $\Lambda$ is quite close to the above result
\begin{equation}
z \ > \ \frac{z_1}{2} \ \Rightarrow \
\frac{m_{t} ^{\mbox{\footnotesize ren.}}}{\Lambda} \ \stackrel{>}{\sim}
\ 0.241 \ (0.237)
\ \Rightarrow \ \Lambda \ \stackrel{<}{\sim} \ 0.75 \mbox{ TeV} \
(0.76 \mbox{ TeV}) \ .
\label{phy3}
\end{equation}

Other authors~\cite{hands} have studied next-to-leading order corrections
in the $1/N_{\mbox{\footnotesize c}}$-expansion
to the gap equation in four dimensions.
They have considered only the two-loop contributions, which by
themselves do not represent all the next-to-leading
order contributions. They did not obtain an upper bound for $\Lambda$.
The authors of~\cite{hands}
dealt mainly with the question of renormalizability
of the model, i.e., of the viability of the limit $\Lambda \to \infty$.
The related question of whether the quadratic divergences cancel exactly
in the TSM has recently been investigated~\cite{blum}, with a goal
of selecting $\Lambda \gg {\cal{O}}(\mbox{TeV})$.

Dealing with the cut-off $\Lambda$ as a physical finite quantity was
crucial in the present paper. We used for the integrals
in the non-QCD sector (e.g.~in calculating ${\cal {J}}_{H} ({\bar p}^2,
{\varepsilon}^2)$) the simple spherical cut-off,
corresponding to the following regions of integration for the
diagrams of Fig.~3 and Figs.~5: $|\bar p|, |{\bar k}^{(1)}|, \ldots,
|{\bar k}^{(\ell)}| \leq \Lambda$. It is a general fact that changing the
cut-off prescription (e.g.~using Pauli-Villars cut-offs in ${\cal {J}}_{H}$),
or imposing the simple
cut-off on some other combinations of the momenta (e.g.~$|{\bar p} +
{\bar k}^{(j)}| \leq \Lambda$, instead of $|{\bar k}^{(j)}| \leq
\Lambda$), would modify numerical results~\cite{gher}. This would affect
quantitatively also the numerical results presented here. However, the main
qualitative features (e.g.~$\Lambda = {\cal{O}}(\mbox{TeV})$) seem to
remain independent of the details of the cut-off prescription. This is
suggested also by the fact that the calculated QCD contributions
(which are numerically less important than those of the non-QCD,
i.e., the scalar contributions) are only mildly dependent of whether
we use for ${\cal {J}}_{\mbox{\footnotesize gl}}
({\bar p}^2,{\varepsilon}^2)$ the ``proper
time'' cut-off (\ref{QCDpt})
or the simple Pauli-Villars subtraction (\ref{QCDPV}).

The effects of the loops of massive electroweak gauge bosons (i.e., their
transverse and longitudinal degrees of freedom), the scalar contributions
of even higher order in $1/N_{\mbox{\footnotesize c}}$
($N_{\mbox{\footnotesize c}} = 3$) ,
as well as dynamical effects of the new physics beyond this relatively
low $\Lambda$, have not been investigated here.
The present results imply that QCD is playing a subordinate role in
the condensation mechanism, and suggest that in the Top-mode Standard
Model (TSM) there are the following two possibilities:
\begin{itemize}
\item The $1/N_{\mbox{\footnotesize c}}$-expansion of the
contributions from the scalar sector~\footnote{
Here we keep QCD contributions aside; they are smaller and can
apparently be organized into a convergent or at least asymptotic
loop series in powers of $\alpha_s(m_t)$; cf.~last paragraph of Section 3.}
to $V_{\mbox{\footnotesize eff}}$ yields a convergent or at least an
asymptotic series, for $N_{\mbox{\footnotesize c}} = 3$. In this case, the
next-to-leading order does not drastically change the gap equation.
Consequently, as shown in the present paper, in such a case the decomposition
of the $\langle \bar t t \rangle$ condensate into the constituent quarks
should occur at a relatively {\it low} energy
$\Lambda ={\cal{O}} (1\mbox{TeV})$.
\item The second possibility occurs when
$\Lambda > {\cal {O}}(1 \mbox{ TeV})$. Then
$1/N_{\mbox{\footnotesize c}}$-expansion (for
$N_{\mbox{\footnotesize c}}=3$) mentioned above is
neither convergent nor asymptotic, and already the next-to-leading order
drastically changes the gap equation. Consequently, the
leading order approximation loses its significance.
We then appear to be in a strongly interacting region -- the
$\langle {\bar t} t \rangle$ condensates appear to interact strongly.
The traditional methods of investigation
(e.g.~the method of calculating $V_{\mbox{\footnotesize eff}}$,
tadpole approach, etc.) are insufficient in this case.
\end{itemize}

\section*{Acknowledgment}
This work was supported in part by the
Deutsche Forschungsgemeinschaft and in part by the European Union
Science Project SC1-CT91-0729.
E.A.P.~wishes to thank W.A.~Bardeen, N.D.V.~to E.G.~Floratos,
and G.C.~to Y.-L.~Wu for illuminating discussions.

\begin{appendix}
\section[]{Tracing in the momentum space}
\setcounter{equation}{0}

In the 4-dimensional Euclidean space we have the following
coordinates and momenta: $\bar x = (i x^0,x^j)$;
$\bar q = (-iq^0,-q^j)$. We denote by $|{\bar x} \rangle$ and
$|{\bar q} \rangle$ the normalized coordinate and momentum
eigenstates. Then the following relations hold:
\begin{displaymath}
\langle \bar{x^{\prime}} | \bar x \rangle = \delta^{(4)}(\bar x -
\bar{x^{\prime}}) \ , \quad
\langle \bar{q^{\prime}} | \bar q \rangle = \delta^{(4)}(\bar q -
\bar{q^{\prime}}) \ , \quad \langle \bar x | \bar q \rangle =
(2 \pi)^{-2} e^{-i \bar x \cdot \bar q} \ .
\end{displaymath}
\begin{equation}
 \int d^4 \bar x | \bar x \rangle \langle \bar x | =
 \int d^4 \bar q | \bar q \rangle \langle \bar q | = 1 \ .
\label{compl}
\end{equation}
The trace of an operator $\hat C$ can then easily be rewritten in
the momentum space
\begin{equation}
Tr \hat C \left(= \int d^4 \bar x \langle \bar x | \hat C | \bar x
\rangle \ \right)
= \int d^4 \bar q \langle \bar q | \hat C | \bar q \rangle \ .
\label{tr2}
\end{equation}
For many operators it is easier to work in the momentum space than
in the $\bar x$-space - e.g.~the operator $\ln {\hat{\cal{A}}}$, if
${\hat{\cal{A}}}$
is translationally invariant and non-diagonal in the $\bar x$-space.
Then the operator $\ln {\hat{\cal{A}}}$  is diagonal in the
momentum space and easier to work with in this space.
When calculating the trace of such operators, we use (\ref{tr2})
in the momentum basis.
All those operators ${\hat C}$ which are translationally
invariant in the $\bar x$-space are diagonal in the momentum space
\begin{equation}
\langle \bar{x^{\prime}} | \hat C | \bar x \rangle =
\langle \bar 0 | \hat C | \bar x - \bar{x^{\prime}} \rangle
\ \  \Rightarrow \ \ \langle \bar{q^{\prime}} | \hat C | \bar q \rangle
= \delta^{(4)}(\bar q - \bar{q^{\prime}} ) \tilde C (\bar q ) \ ,
\ \mbox{where: }
 \tilde C (\bar q) = \int d^4 \bar x e^{-i \bar q \cdot \bar x}
\langle \bar 0 | \hat C | \bar x \rangle \ .
\label{trinv}
\end{equation}
The trace of such an operator can be obtained from (\ref{tr2}),
once we know the Fourier-transform $\tilde C (\bar q)$
\begin{equation}
Tr \hat C = (2 \pi)^{-4} \Omega \int d^4 \bar q \tilde C(\bar q) \ ,
\quad \mbox{where: } \ \Omega = (2\pi)^4 \lim_{\bar q \to 0}
\delta^{(4)}(\bar q) = \int d^4 \bar x \ .
\label{tr3}
\end{equation}
It is straightforward to check that for any power of such an operator
$\hat C$ we have
\begin{equation}
\langle \bar{q^{\prime}} | \hat C^n | \bar q \rangle =
\delta^{(4)}(\bar q - \bar{q^{\prime}} ) (\tilde C (\bar q))^n
\qquad (n=0, \pm 1, \pm 2, \ldots ) \ .
\label{pow}
\end{equation}
The operator $\hat B [{\sigma}_0]$, as defined by (\ref{hatB}) and
(\ref{omega}), is translationally invariant. Therefore
\begin{displaymath}
\langle \bar{q^{\prime}}; j,a | \hat B [{\sigma}_0] |
\bar q ; k,b \rangle  =
\delta^{(4)}(\bar q - \bar{q^{\prime}}) \tilde B_0(\bar q; j,k;a,b) \ ,
\end{displaymath}
\begin{equation}
\tilde B_0 (\bar q; j,k;a,b)  =  \delta_{ab} \delta_{jk}
 (- \bar q \llap / + \delta_{j1} {\lambda} {\sigma}_0 ) \ ,
\ \ \mbox{where: }
{{\bar q} \llap /} = {\bar \gamma}^{\mu} {\bar q}^{\mu}  \
({\bar \gamma}^0 = i {\gamma}^0, \ {\bar \gamma}^j = {\gamma}^j).
\label{B0}
\end{equation}
In order to calculate $Tr(\ln \hat B [{\sigma}_0])$ of eq.~(\ref{1loop1}),
we note first that $\ln \hat B[{\sigma}_0]$ can be
written as a Taylor series in powers of $(\hat B[{\sigma}_0]-1)$.
Applying to these powers relation (\ref{pow}), we obtain
\begin{equation}
\langle \bar{q^{\prime}}; j,a | \ln \hat B [{\sigma}_0] |
\bar q ; k,b \rangle =
\delta^{(4)}(\bar q - \bar{q^{\prime}}) \delta_{ab} \delta_{jk}
\ln (- \bar q \llap / + \delta_{j1} {\lambda} {\sigma}_0 ) \ .
\label{lnB0}
\end{equation}
Applying now the relations (\ref{tr2}) and (\ref{tr3}), we obtain
$Tr(\ln \hat B [{\sigma}_0])$
\begin{equation}
Tr \ln \hat B [{\sigma}_0]  =
 =  (2 \pi)^{-4} \Omega N_{\mbox{\footnotesize c}}
 tr_{\mbox{\scriptsize f}} \int d^4 \bar q
 \ln \left( - \bar q \llap / + {\lambda} {\sigma}_0 \right) + \cdots \ ,
\label{trlnB0}
\end{equation}
where the dots represent an ${\sigma}_0$-independent term (corresponding
to the isospin index $j=2$), which
is irrelevant for $V^{(1)}_{\mbox{\footnotesize eff}} ({\sigma}_0)$.
The factor $N_{\mbox{\footnotesize c}}$ (the number of colors)
comes from summing over
the color indices $a$, $b$.
(\ref{trlnB0}) gives us the result on the r.h.s.~of eq.~(\ref{1loop1})
(replace: ${\bar q} \mapsto -{\bar k}$).

Next we derive the result of eq.~(\ref{Aomega}). The operator
$\hat A$, which was introduced in eq.~(\ref{pathomega}), was defined
through the relation (\ref{Axbas}), or equivalently
(see eqs.~(\ref{pathint4}) and (\ref{pathomega}))
\begin{equation}
2 \int d^4 \bar x {\sigma}_1(\bar x)^2  + {\lambda}^2
Tr\lbrack (\hat B[{\sigma}_0]^{-1} {\hat {\sigma}_1} )^2 \rbrack =
\int d^4 \bar x d^4 \bar{x^{\prime}} {\sigma}_1(\bar{x^{\prime}})
\langle \bar{x^{\prime}} | \hat A | \bar x \rangle {\sigma}_1(\bar x) \ .
\label{relom}
\end{equation}
In order to obtain the expression for the matrix element $\langle
\bar{x^{\prime}} | \hat A | \bar x \rangle$, we must calculate
$Tr \lbrack (\hat B[{\sigma}_0]^{-1} {\hat {\sigma}_1} )^2 \rbrack$.
First, we use (\ref{tr2}) and repeatedly apply the completeness insertions
(\ref{compl})
\begin{eqnarray}
\lefteqn{ Tr \lbrack (\hat B[{\sigma}_0]^{-1} {\hat {\sigma}_1} )^2 \rbrack
= \langle \bar q | \hat B[{\sigma}_0]^{-1} | \bar q^{(1)} \rangle
\langle \bar q^{(1)} | \bar x^{(1)} \rangle
\langle \bar x^{(1)} | {\hat {\sigma}_1} | \bar x^{(2)} \rangle
\langle \bar x^{(2)} | \bar q^{(2)} \rangle \times}
\nonumber \\
& & \times \langle \bar q^{(2)} | \hat B[{\sigma}_0]^{-1} | \bar q^{(3)}
\rangle  \langle \bar q^{(3)} | \bar x^{(3)} \rangle
\langle \bar x^{(3)} | {\hat {\sigma}_1} | \bar x^{(4)} \rangle
\langle \bar x^{(4)} | \bar q \rangle \ .
\label{insert}
\end{eqnarray}
Here, we implicitly assume integrations over all momenta ($\bar q$,
$\bar q^{(1)}$, $\ldots$) and coordinates
($\bar x^{(1)}$, $\bar x^{(2)}$, $\ldots$). Furthermore, the color,
isospin and spinor degrees of freedom were omitted in the notation
of (\ref{insert}); summation over these indices and tracing over spinor
degrees of freedom ($tr_{\mbox{\scriptsize f}}$) is implicitly assumed.
The matrix elements of $\hat B[{\sigma}_0]^{-1}$, which appear on the
r.h.s.~of (\ref{insert}), are obtained from (\ref{B0}) and (\ref{pow})
(for $n=-1$). Using in addition (\ref{compl}) for
$\langle \bar x | \bar q \rangle$, and the locality of the quantum
fluctuation operator ${\hat {\sigma}_1}$ (i.e.,
$\langle \bar{x^{\prime}} | {\hat {\sigma}_1} | \bar x \rangle =
{\sigma}_1(\bar x) \delta^{(4)}(\bar x - \bar{x^{\prime}})$),
we obtain from (\ref{insert}) after some bookkeeping
\begin{equation}
Tr \lbrack (\hat B[{\sigma}_0]^{-1} {\hat {\sigma}_1} )^2 \rbrack =
N_{\mbox{\footnotesize c}} tr_{\mbox{\scriptsize f}}
\int d^4 \bar x d^4 \bar{x^{\prime}}
{\sigma}_1(\bar{x^{\prime}}) {\sigma}_1(\bar x)
\int  \frac{ d^4\bar q d^4\bar{q^{\prime}} } {(2\pi)^8}
\frac{ \exp \left[ i \left( \bar q - \bar{q^{\prime}} \right)
\cdot \left( \bar x - \bar{x^{\prime}} \right) \right] }
{ (- {{\bar q} \llap /} + {\lambda} {\sigma}_0 )
  (- {{\bar q} \llap /}^{\prime} + {\lambda} {\sigma}_0 ) }
+ \cdots \ .
\label{trBo2}
\end{equation}
Here, the factor $N_{\mbox{\footnotesize c}}$ came from tracing
over the color degrees of freedom. Dots represent a term independent
of ${\sigma}_0$ (corresponding to the isospin indices $j=k=2$), i.e., this
term is irrelevant for $V_{\mbox{\footnotesize eff}}({\sigma}_0)$
(cf.~eq.~(\ref{pathint4})).
Performing in (\ref{trBo2}) the replacements $\bar{q^{\prime}} \to
{\bar k}$, ${\bar q} \to  {\bar p} + {\bar k}$, combining the obtained
expression with the relation (\ref{relom}), and introducing a physical
energy cut-off $(\bar k^2)_{max} = \Lambda^2$, we immediately
obtain the result of eq.~(\ref{Aomega}), i.e., the
explicit expression for the matrix element  $\langle \bar{x^{\prime}}
|\hat A| \bar x \rangle$, its Fourier transform $\tilde A (\bar p)$
and the related function ${\cal{K}}_{H}  (\bar p^2, \lambda^2 {\sigma}_0^2)$.
Note that the operator $\hat A$ is the
kernel for the quadratic quantum fluctuations of the Higgs field
around the minimum.

Finally, according to eq.~(\ref{res1}), $Tr \ln \hat A$ remains to be
calculated
in order to obtain $V^{\mbox{\scriptsize (ntl)}} _{\mbox{\footnotesize eff}}$.
This can be done now, by
using the same approach as in (\ref{trlnB0}) for calculating
$Tr \ln \hat B[{\sigma}_0]$
\begin{equation}
Tr \ln \hat A  =
 (2 \pi)^{-4} \Omega \int d^4 \bar p \ln \tilde A(\bar p)
 =  (2 \pi)^{-4} \Omega \int d^4 \bar p \ln \lbrack 1
- 2 {\lambda}^2 N_{\mbox{\footnotesize c}} {\cal{K}}_{H}
(\bar p^2, \lambda^2 {\sigma}_0^2) \rbrack + \cdots \ ,
\label{trlnA}
\end{equation}
where the dots represent an irrelevant ${\sigma}_0$-independent term.
This time the trace is performed only over the 4-momentum space,
because the operator $\hat A$ (eq.~(\ref{Aomega})) has no color, isospin
or spinor degrees of freedom any more.
Using the relation $d^4 \bar p = \pi^2 d \bar p^2 \bar p^2$
(the integrand in
(\ref{trlnA}) is spherically symmetric), introducing a physical
energy cut-off $(\bar p^2)_{max}= \Lambda^2$, and
rescaling the momentum $\bar p^2 \mapsto \Lambda^2 \bar p^2$,
leads directly to the final result (\ref{res2}) and (\ref{Jdef}).
This is the next-to-leading order (beyond 1-loop) contribution
of the composite Higgs sector to the effective potential.

\section[]{J functions for the effective potential}
\setcounter{equation}{0}

The function ${\cal{J}}_{H}  (\bar p^2, \varepsilon^2)$, which is
needed for calculating the $V^{\mbox{\scriptsize (ntl)}}
_{\mbox{\footnotesize eff}} $ in eq.~(\ref{res2}) and is given in
an integral form in eq.~(\ref{Jdef}), can be calculated analytically.
In the 4-dimensional Euclidean space, we can write
\begin{equation}
\bar p \cdot \bar k = |\bar p| |\bar k| \cos \theta \ ,
\quad d^4 \bar k = (2 \pi) |\bar k|^2 \sin^2 \theta d |\bar k|^2 d \theta
\qquad (0 \leq \theta \leq \pi) \ .
\end{equation}
The integration over $\theta$ in (\ref{Jdef}) can be performed by using
(cf.~Ryzhik and Gradshteyn, 3.644/4)
\begin{equation}
\int_0^{\pi} \frac{d \theta \sin^2\theta}{({\cal{A}}
+ {\cal{B}} \cos \theta)}
= \pi \frac{{\cal{A}} }{{\cal{B}}^2} \left[ 1 -
\sqrt{(1-\frac{{\cal{B}}^2}{{\cal{A}}^2})} \right] \ .
\label{rgang}
\end{equation}
Subsequently, the radial integration over $|\bar k|^2$ in (\ref{Jdef})
is performed by using well-known formulas (e.g.~Ryzhik and Gradshteyn,
2.261, 2.266, 2.267). It leads to the result
\begin{eqnarray}
\lefteqn{ {\cal{J}}_{H}  (\bar p^2, {\varepsilon}^2) =
  {\Big \{} \frac{3}{4} -
  \frac{1}{2}{\varepsilon}^2 \ln({\varepsilon}^{-2}+1)
  +  \frac{ \left( 1+{\varepsilon}^2 \right)^2 }{ 8 {\bar p}^2 }
  - \frac{{\bar p}^2}{8}
  - \frac{(1-\bar p^2 + {\varepsilon}^2)}{8 \bar p^2} {\cal{B}}
- \frac{{\varepsilon}^2}{2} \ln \left( \frac{a_3}{2 {\varepsilon}^2} \right) }
 \nonumber\\
 & & - \frac{1}{4} (\bar p^2 + 4 {\varepsilon}^2) \left[ 1
 + \frac{(1+{\varepsilon}^2-{\cal{B}} )}{
\bar p^2} +(1- {\cal{A}} ) \ln ({\varepsilon}^{-2}+1)
+ {\cal{A}} \ln \left( \frac{a_1}{a_2} \right)
+ \ln \left( \frac{a_3}{2 {\varepsilon}^2} \right) \right]  {\Big \}} \ ,
\label{J}
\end{eqnarray}
\begin{displaymath}
\mbox{where: } \qquad
{\cal{A}}  =  \sqrt{1+ 4 \frac{{\varepsilon}^2}{\bar p^2}} \ , \qquad
{\cal{B}} = \sqrt{(1-\bar p^2+{\varepsilon}^2)^2
+ 4 \bar p^2 {\varepsilon}^2} \ ,
\end{displaymath}
\begin{equation}
a_1 =  (\bar p^2 + 3 {\varepsilon}^2 - 1 + {\cal{A}} {\cal{B}} ) \ , \ \
a_2= \bar p^2 + 3 {\varepsilon}^2 + (\bar p^2 + {\varepsilon}^2) {\cal{A}}
\  , \ \ a_3 = 1 - \bar p^2 + {\varepsilon}^2 + {\cal{B}} \ .
\label{Jadd}
\end{equation}
The partial derivative
$\partial {\cal{J}}_{H}  / \partial {\varepsilon}^2$,
which appears in an integrand of the gap equation
(\ref{2lgap}), is
\begin{eqnarray}
\lefteqn{ \frac{\partial {\cal{J}}_{H}  (\bar p^2, {\varepsilon}^2)}{\partial
{\varepsilon}^2} =
 {\Big \{ } \frac{(\bar p^2+ 6 {\varepsilon}^2)}{4 {\varepsilon}^2
 (1+{\varepsilon}^2)}
-\frac{(\bar p^2 + {\varepsilon}^2)(\bar p^2 + 7 {\varepsilon}^2)}{4
\bar p^2 {\varepsilon}^2} -\frac{3}{4 \bar p^2} +
\frac{{\cal{B}} (3+\bar p^2 + 7 {\varepsilon}^2)}{4 \bar p^2
(1+{\varepsilon}^2)} }
 \nonumber\\
 & &   -\frac{3}{2} \ln \left( \frac{a_3}{2} \right)
- \frac{3}{2}{\cal{A}} \ln \left( \frac{a_1}{a_2} \right)
+\frac{3}{2} ({\cal{A}} - 1) \ln(1 + {\varepsilon}^2) +
\frac{3}{2} (2- {\cal{A}} ) \ln {\varepsilon}^2
  {\Big \} } \ ,
\label{derivJ}
\end{eqnarray}
where the notations of (\ref{Jadd}) are used. This expression can be
obtained by the same method as (\ref{J}), or by direct differentiation
of (\ref{J}).
A form of ${\cal{J}}_{H}  (\bar p^2, \varepsilon^2)$ accurate up to
${\cal{O}}({\varepsilon}^4 \ln {\varepsilon}^2)$ (note:
${\bar p}^2, \ {\varepsilon}^2 \ \propto \ 1/\Lambda^2$) is
\begin{eqnarray}
\lefteqn{ {\cal{J}}_{H}  (\bar p^2, {\varepsilon}^2) =
 {\Big \{ } \left( 1- 2 {\varepsilon}^2
  + 3 {\varepsilon}^2 \ln{\varepsilon}^2 \right) +
 {\bar p}^2 \left( - \frac{3}{4} + \frac{1}{2} \ln {\varepsilon}^2
   - \frac{1}{2} {\varepsilon}^2 \right) - }
     \nonumber\\
& &
- \frac{1}{2} {\bar p}^2
 \left( 1 + \frac{4 {\varepsilon}^2 }{ {\bar p}^2 } \right)^{3/2}
  \ln \left[ 1 + \frac{ {\bar p}^2 }{2 {\varepsilon}^2 } -
  \sqrt{ \frac{ {\bar p}^2 }{ {\varepsilon}^2 }
  \left( 1+\frac{ {\bar p}^2 }{ 4{\varepsilon}^2 } \right) }
    \right] + {\cal{O}}
    \left( {\varepsilon}^4 \ln {\varepsilon}^2 \right) {\Big \} } \ .
\label{Jexpan}
\end{eqnarray}
Furthermore, the ${\cal{J}}_{H}
(\bar p^2 = 0, {\varepsilon}^2)$ can be obtained either
from (\ref{J}), or by direct spherical integration
\begin{equation}
{\cal{J}}_{H}  (\bar p^2 = 0, {\varepsilon}^2) =
\left[ 1 - 3 {\varepsilon}^2 \ln ({\varepsilon}^{-2}+1) +
\frac{2 {\varepsilon}^2}{1+{\varepsilon}^2} \right] \ .
\label{Jp0}
\end{equation}
The expression for $\alpha$ (eq.~(\ref{Jint2})) was assumed to be
positive, because otherwise the path integral derivation of
$V^{\mbox{\scriptsize (ntl)}} _{\mbox{\footnotesize eff}} $ presented
there would not be mathematically justifiable
\begin{equation}
\alpha = \frac{\Omega}{\tilde A (\bar p = 0,\lambda^2\sigma_0^2)}
 \ > \ 0 \quad \Rightarrow \quad a J (\bar p^2=0, {\varepsilon}^2) \ < \ 1
\quad (a= \frac{G N_{\mbox{\footnotesize c}} \Lambda^2}{8 \pi^2} \sim 1) \ .
\label{alpos1}
\end{equation}
It can be checked numerically that for ($0 \leq {\bar
p}^2, {\varepsilon}^2 \leq 1$) the function ${\cal{J}}_{H} (\bar p^2,
{\varepsilon}^2)$ has the following properties:
${\cal{J}}_{H} ({\bar p}^2,{\varepsilon}^2) \leq 1$, where the equality
holds only at ${\bar p}^2={\varepsilon}^2=0$; ${\cal{J}}_{H}
({\bar p}^2,{\varepsilon}^2)$ is a monotonously decreasing function of
${\bar p}^2$ if ${\varepsilon}^2 < 0.63$.
Hence, (\ref{alpos1}) implies the general condition
\begin{equation}
a {\cal{J}}_{H}  (\bar p^2, {\varepsilon}^2) \ < \ 1 \qquad
\mbox{for all: } \ (0 \leq \bar p^2 \leq 1) \ .
\label{nosing}
\end{equation}
However, this is precisely the condition that the expression
$V_{\mbox{\footnotesize eff}}^{\mbox{\scriptsize (ntl)}} $ in
eq.~(\ref{res2}) does not contain any
singularities in the integrand. Therefore, the condition $\alpha > 0$,
which was needed so that the derivation of $V^{\mbox{\scriptsize (ntl)}}
_{\mbox{\footnotesize eff}} $ by path integral method
was mathematically consistent, guarantees also that the obtained result is
mathematically well-behaved. From the diagrammatic point of view
(Appendix C), the condition (\ref{nosing}) means that the diagrammatic
sum of the next-to-leading order terms in the
$1/N_{\mbox{\footnotesize c}}$-expansion contributing to the effective
potential (i.e., $V^{\mbox{\scriptsize (ntl)}} _{\mbox{\footnotesize eff}}$)
does not diverge (cf.~eqs.~(\ref{Vl3}), (\ref{resC})), and in fact leads
to the logarithm in the integrand of eq.~(\ref{res2}).
The ``non-singularity'' condition (\ref{nosing}) holds safely for all
such ${\varepsilon}^2$ which satisfy
\begin{equation}
{\varepsilon}^2   \ > \  0.25 z_1 \qquad
   \left( z_1= \left(  m_{t}^{(1)}/\Lambda \right)^2 \right) \ ,
\label{alpos4}
\end{equation}
where $z_1$ is the solution of the $(0+1)$-loop gap equation.
The mentioned singularities appear in Figs.~1a-1e as poles
of $V_{\mbox{\footnotesize eff}}({\varepsilon}^2)$ (full lines) in
these regions of small ${\varepsilon}^2$
-- features which are not physical, but represent some artifacts
of the mathematical approach of the present paper -- namely, the
$1/N_{\mbox{\footnotesize c}}$-expansion in the electroweak sector. In
any way, if the calculated $1/N_{\mbox{\footnotesize c}}$-corrections to
$V_{\mbox{\footnotesize eff}}$ were to change
the position of the minimum $z$ as drastically as by factor $4$ or
more, i.e., if singularities were to appear near the new minimum
(cf.~(\ref{alpos4})), then the expansion of the electroweak
contributions to $V_{\mbox{\footnotesize eff}}$ in powers of
$1/N_{\mbox{\footnotesize c}}$ would be a
highly divergent and senseless series, anyway.
Since these singularities always lie safely away from the minima of
the effective potential (${\varepsilon}^2 \approx z$) given in
Table 1, they do not influence
the conclusions about the value and the existence of this minimum
in any appreciable way, and are hence irrelevant for the purpose of the
present paper. In Figs.~1a-1e we have not continued the (full) lines
of $V_{\mbox{\footnotesize eff}}$ into these singular
small-${\varepsilon}^2$ regions.

The integrals (\ref{QCD3})-(\ref{QCDpt}), connected with the dominant
gluonic contribution to $V_{\mbox{\footnotesize eff}}$, have also
explicit solutions. The
resulting function ${\cal {J}}_{\mbox{\footnotesize gl}} $
(cf.~(\ref{QCD2})-(\ref{QCDresdef}))
is in the case of one simple Pauli-Villars cut-off
\begin{eqnarray}
{\cal {J}}_{\mbox{\footnotesize gl}} ^{(\mbox{\scriptsize P.V.})}
\left({\bar p}^2, {\varepsilon}^2 \right) & = &
\int_0^1 dz z (1-z) \ln \left[ \frac{z(1-z){\bar p}^2 +
{\varepsilon}^2}{z(1-z){\bar p}^2+1} \right]
 \nonumber\\
 &= & -\frac{1}{6} \left(2 \frac{{\varepsilon}^2}{{\bar p}^2} - 1 \right)
 {\cal {D}}\left( \frac{{\varepsilon}^2}{{\bar p}^2} \right)
 +\frac{1}{6} \ln {\varepsilon}^2
 +\frac{1}{6} \left(2 \frac{1}{{\bar p}^2} - 1 \right)
 {\cal {D}}\left( \frac{1}{{\bar p}^2} \right) \ ,
 \label{JglPV}
 \end{eqnarray}
 and in the case of the ``proper time'' cut-off
 \begin{eqnarray}
 {\cal {J}}_{\mbox{\footnotesize gl}} ^{(\mbox{\footnotesize p.t.})}
 \left({\bar p}^2, {\varepsilon}^2 \right) & = &
 -\frac{1}{6} \left(2 \frac{{\varepsilon}^2}{{\bar p}^2} - 1 \right)
 {\cal {D}}\left( \frac{{\varepsilon}^2}{{\bar p}^2} \right)
 +\frac{1}{6} \ln {\varepsilon}^2 + \frac{2}{9}
 \nonumber\\
 & & -\frac{1}{6} \left( \frac{{\bar p}^2}{5}+{\varepsilon}^2 \right)
 +\frac{1}{4}\left( \frac{{\bar p}^4}{140}+
 \frac{{\bar p}^2 {\varepsilon}^2}{15}+\frac{{\varepsilon}^4}{6} \right)+
 {\cal {O}}\left( {\bar p}^6, {\varepsilon}^6 \right) \ .
 \label{Jglpt}
 \end{eqnarray}
 Here we denoted by ${\cal {D}}$ the integral
 \begin{equation}
 {\cal {D}}\left( w \right) = \int_0^1 dz \ln \left[1 + \frac{z(1-z)}{w}
 \right] = -2 + \sqrt{\left( 4 w+1 \right)} \ln \left[ \frac{
 \sqrt{ \left( 4w+1 \right) }+1}{\sqrt{ \left( 4w+1 \right) }-1}
 \right] \ .
 \label{defD}
 \end{equation}
 The corresponding derivatives
 $\partial {\cal {J}}_{\mbox{\footnotesize gl}}
 ( {\bar p}^2, \varepsilon^2 )/\partial \varepsilon^2$
 are obtained in a straightforward manner from the above expressions,
 by using also the following integral
 \begin{equation}
 \frac{d {\cal {D}} \left( w \right) }{dw} =
 -\frac{1}{w}+ \frac{2}{\sqrt{ \left( 4w+1 \right) }} \ln \left[ \frac{
 \sqrt{ \left( 4w+1 \right) }+1}{\sqrt{ \left( 4w+1 \right) }-1} \right] \ .
 \end{equation}
 We note that, unlike ${\cal {J}}_{H} $, the gluonic function
 ${\cal {J}}_{\mbox{\footnotesize gl}} $
 is negative in the entire region of ${\bar p}^2$ and ${\varepsilon}^2$
 of our interest, and therefore does not lead to any possible singularities
 in the logarithm of $V_{\mbox{\footnotesize eff}}
 ^{\mbox{\scriptsize (gl)}} $ in (\ref{QCDres}).

\section[]{Diagrammatic derivation of $V^{\mbox{\scriptsize (ntl)}}
_{\mbox{\footnotesize eff}} $; mass renormalization}
\setcounter{equation}{0}

In this Appendix we show that
$V^{\mbox{\scriptsize (ntl)}} _{\mbox{\footnotesize eff}} $ of
eq.~(\ref{res2}) can be rederived by a diagrammatic method and, in addition,
that it represents precisely all the effects of loops which are
next-to-leading order in the formal
$1/N_{\mbox{\footnotesize c}}$-expansion (these are terms
beyond one loop). It is evident from eq.~(\ref{1lgap}) that the
4-fermion coupling $G$ is of order $1/N_{\mbox{\footnotesize c}}$ (for a
fixed $\Lambda^2$). This fact will be used repeatedly throughout this
Appendix.

We will use the original notations of the Lagrangian (\ref{TSM1}),
in order to see explicitly that the dependence of the results on
the undetermined bare mass $M_0$ drops out. Furthermore, we will
introduce $g = M_0 \sqrt{G}$ which is dimensionless.
We employ the usual diagrammatic method for
$V_{\mbox{\footnotesize eff}}$~\cite{bailin},
which has been applied in ref.~\cite{gceap} to the 1-loop graphs.
\begin{equation}
V_{\mbox{\footnotesize eff}}(H_0) = i \sum_{m=1}^{\infty} \frac{1}{(2m)!}
 \tilde \Gamma_{H} ^{(2m)}(p_1, \ldots , p_{2m}) {\Big|}_{(\lbrace p_k
 \rbrace= \lbrace 0 \rbrace)} H^{(2m)}_0
\ .
\label{Veffgraph}
\end{equation}
Here, $\tilde \Gamma_{H} ^{(2m)}$'s are 1-PI Green functions corresponding
to the diagrams of the Lagrangian (\ref{TSM1}) with $2m$ outer
legs of the (yet non-dynamical) Higgs $H_0$ fields with zero
momenta\footnote{
It will be shown later in this Appendix that diagrams with an odd number
of $H_0$-outer legs do not contribute.}.
The 1-loop diagrams contributing
to $\tilde \Gamma_{H} ^{(2m)}$'s are depicted in Figs.~2a-2c, and they
lead to $V^{(1)}_{\mbox{\footnotesize eff}} $ of
eq.~(\ref{1loop2})~\cite{gceap}. The
$(\ell+1)$-loop 1-PI diagrams contributing to $\tilde \Gamma^{(2m)}_{H} $
are depicted in Fig.~3. We note that {\it no} other $(\ell+1)$-loop 1-PI
diagrams are next-to-leading order in the formal
$1/N_{\mbox{\footnotesize c}}$-expansion.
For example, the 3-loop and 4-loop 1-PI diagrams of Figs.~4a-4b
(when the external legs are cut away) are not proportional to
$(G^2 N_{\mbox{\footnotesize c}}^2)$, $(G^3 N_{\mbox{\footnotesize c}}^3)$
($= {\cal{O}} (1)$), but rather to
$(G^2 N_{\mbox{\footnotesize c}})$, $(G^3 N_{\mbox{\footnotesize c}}^2)$
($ = {\cal{O}} (1/N_{\mbox{\footnotesize c}})$),
respectively. On the other hand, the diagrams in Fig.~3
(when the external legs are cut out) are proportional to ${\cal{O}}(1)$,
i.e., they are next-to-leading order in the
$1/N_{\mbox{\footnotesize c}}$-expansion.
We can understand this by counting the number of quark loops
and of Yukawa couplings. In Fig.~3, the total number of loops is
$(\ell + 1)$, and the number of (top) quark loops among them
is $n_{q}=\ell$. In all other 1-PI diagrams with the total number
of loops $(\ell+1)$ (e.g.~those in Figs.~4a-4b) is the number of
(top) quark loops less than $\ell$: $n_{q} < \ell$.
On the other hand, the number of Yukawa couplings is $2 \ell$ in all
1-PI diagrams with the total number of loops $(\ell+1)$, and therefore
the Yukawa couplings yield a factor $(\sqrt{G})^{2 \ell}$.
Each quark loop gives a factor $N_{\mbox{\footnotesize c}}$ by tracing
over the color degrees of freedom. The $2m$ external Higgs lines
give the factor $(g H_0)^{2m}$ ($g= \sqrt{G} M_0$)~\footnote{
The parameter $\varepsilon^2 = (g H_0)^2/(2 \Lambda^2)$
(cf.~eq.~(\ref{Jdef})) turns out to be
the natural variable of ${\cal{O}}(1/N_{\mbox{\footnotesize c}}^0)$
in $V_{\mbox{\footnotesize eff}}$.}. In total, this
gives contributions to $V_{\mbox{\footnotesize eff}}$ of
(\ref{Veffgraph}) proportional to
$(G^{\ell} N_{\mbox{\footnotesize c}}^{n_{q}})(g H_0)^{2m} =
{\cal{O}}(1/N_{\mbox{\footnotesize c}}^{\ell-n_{q}})
(g H_0)^{2m}$. In the case of the diagrams of Fig.~3,
${\cal{O}}(1/N_{\mbox{\footnotesize c}}^{\ell-n_{q}})$ is ${\cal{O}}(1)$,
in the case of other diagrams (e.g.~those of Figs.~4a-4b)
we have ${\cal{O}}(1/N_{\mbox{\footnotesize c}}^{\ell-n_
{\mbox{\footnotesize f}} }) \leq {\cal{O}}(1/N_{\mbox{\footnotesize c}})$.
Thus, all 1-PI diagrams with $(\ell+1)$ loops and $\ell \geq 1$,
other than those of Fig.~3, give contributions to
$V_{\mbox{\footnotesize eff}}(g H_0)$
which are of lower order than the next-to-leading order terms in
in the $1/N_{\mbox{\footnotesize c}}$-expansion.
Therefore, we will ignore all such diagrams.
We note in passing that there is only one 1-PI diagram with two loops, and
it is of the type depicted in Fig.~3, i.e., of the next-to-leading order
in the $1/N_{\mbox{\footnotesize c}}$-expansion.

The 1-PI Green function resulting from the $(\ell+1)$-loop diagrams
with $(2m)$-outer Higgs legs of Fig.~3 can be written as
\begin{equation}
\tilde \Gamma_{H} ^{(2m; \ell+1)} (p_1, \ldots, p_{2m}){\Big|}_{
(\lbrace p_j \rbrace = \lbrace 0 \rbrace)}
 = \sum_{ m_1, \ldots, m_{\ell} \geq 0 \atop m_1+\cdots m_{\ell}=m }
  {\cal{G}}_{m_1 \cdots m_{\ell}} \ .
\label{G1}
\end{equation}
Here, the $(2m)$ external Higgs lines are distributed among the quark loops
as follows: each term ${\cal{G}}_{m_1 \cdots m_{\ell}}$ represents the
contribution of those diagrams of Fig.~3 which have on their $\ell$ loops
of massless (top) quarks $2m_1,\ldots,2m_{\ell}$ outer $H_0$-legs with zero
momenta, respectively. Hence:
\begin{eqnarray}
\lefteqn{
{\cal{G}}_{ m_1 \cdots m_{\ell} \atop (m_1+ \cdots + m_{\ell} = m) } =
\frac{i^{2m+2\ell}}{(2m+2\ell)!} N_{\mbox{\footnotesize c}}^{\ell}
 (-1)^{\ell} \left( -\frac{g}{\sqrt{2}} \right)^{2m+2\ell}
 \int \frac{d^4p d^4k^{(1)} \cdots d^4k^{(\ell)}}{(2\pi)^{4(\ell+1)}}
 \times }
\nonumber\\
& & \left[ \prod_{j=1}^{\ell} tr_{\mbox{\footnotesize f}}
\left( \sum_{n_j=0}^{2m_j}\frac{i^{2m_j+2}}
 { ({k \llap /}^{(j)})^{2m_j+1-n_j} (p \llap / + {k \llap /}^{(j)})^{n_j+1}}
 \right) \right]
\times \left( \frac{i}{-M_0^2} \right)^{\ell} \times
N_{H} ^{\mbox{\scriptsize in}} N_{H} ^{\mbox{\scriptsize out}} N_{q} \ .
\label{G2}
\end{eqnarray}
Here, $k^{(j)}$ and $(p+k^{(j)})$ are the two momenta in the $j^{th}$
(massless top) quark loop; $p$ is the momentum of the $\ell$ internal Higgs
propagators, and each such propagator is simply $(-i/M_0^2)$ (Higgses are
non-dynamical); $(-g/\sqrt{2})$ is the Yukawa coupling (cf.~eq.~(\ref{TSM1});
$g=M_0 \sqrt{G}$); $(-1)^{\ell}$ is the Fermi statistical factor from the
$\ell$ quark-loops; $N_{\mbox{\footnotesize c}}^{\ell}$ is the factor
arising from tracing over the colors of the $\ell$ quark loops;
$N_{H} ^{\mbox{\scriptsize in}}$, $N_{H} ^{\mbox{\scriptsize out}}$,
$N_{q}$ are the numbers of possible contractions between the internal Higgs
fields, between the external Higgs fields, and between the internal (top)
quarks, respectively (in the framework of the formalism of the Wick theorem).
We label each quark loop by the number $j$ ($j=1,\ldots,\ell$).
The $j^{th}$ quark loop has altogether $2 m_j$
external Higgs legs of momentum zero and two legs of internal Higgs
propagators of momentum $p$ attached to it. The external Higgs legs
are distributed on this loop so that there are $n_j$ legs attached to it
outside the chain and $(2 m_j-n_j)$ legs attached to it inside
the chain (see Fig.~3).
When performing the trace, we have to sum over all possible distributions
of the external Higgs lines, i.e., varying $n_j$ from $0$ to $2 m_j$.
Incidentally, it follows from the above expression (\ref{G2})
that the numbers $2m_1, \ldots, 2m_{\ell}$ (and their sum $2m$, i.e., the
number of external legs) must be even, because trace of a product of an odd
number of $\gamma$ matrices is zero.

Next, we must count carefully the contractions
$N_{H}^{\mbox{\scriptsize in}}$, $N_{H} ^{\mbox{\scriptsize out}}$
and $N_{q}$, as dictated by the Wick theorem. We find
\begin{displaymath}
N_{H} ^{\mbox{\scriptsize in}} = { 2m+2\ell \choose 2 }
{ 2m+2\ell-2 \choose 2 } \cdots
 { 2m+2 \choose 2 } \frac{1}{\ell !} =
 \frac{(2m+2\ell)!}{(2m)! (2!)^{\ell} {\ell}!} \ ,
\end{displaymath}
\begin{equation}
N_{H} ^{\mbox{\scriptsize out}} = (2m)! \ , \qquad N_{q} = (2m)!
(\ell - 1)! 2^{\ell - 1} \ .
\label{contr}
\end{equation}
Using (\ref{Veffgraph}), (\ref{G1}) and (\ref{G2}), this gives us
the following expression for those contributions of the
$(\ell + 1)$-loops ($\ell \geq 1$) to the effective potential which
are next-to-leading order in $1/N_{\mbox{\footnotesize c}}$
\begin{eqnarray}
\lefteqn{
V_{\mbox{\footnotesize eff}}^{(\ell+1)-loops}(H_0)  =
\sum_{m=0}^{\infty} \left( \frac{G N_{\mbox{\footnotesize c}}}{2}
\right)^{\ell}
\left( \frac{g H_0}{\sqrt{2}} \right)^{2m}
\frac{i^{\ell + 1}}{2 \ell}
 \int \frac{d^4p d^4k^{(1)} \cdots d^4k^{(\ell)}}{(2\pi)^{4(\ell+1)}}
 \times }
   \nonumber\\
& & \sum_{ m_1,\cdots,m_{\ell} \geq 0 \atop m_1+ \cdots + m_{\ell}=m }
\left[ \prod_{j=1}^{\ell}
tr_{\mbox{\footnotesize f}}  \left( \sum_{n_j=0}^{2m_j}\frac{1}
 { ({k \llap /}^{(j)})^{2m_j+1-n_j} (p \llap / + {k \llap /}^{(j)})^{n_j+1}}
 \right) \right] \ .
\label{Vl1}
\end{eqnarray}
It is straightforward to see that for the expansion in powers of
$(g H_0/\sqrt{2})$ the following crucial identity holds:
\begin{eqnarray}
\lefteqn{
\prod_{j=1}^{\ell} tr_{\mbox{\footnotesize f}}  \left[
\frac{1}{({k \llap /}^{(j)}-\frac{g H_0}{\sqrt{2}})}
\frac{1}{(p \llap / + {k \llap /}^{(j)}-\frac{g H_0}{\sqrt{2}})}
\right]  =  \sum_{m=0}^{\infty} \left( \frac{g H_0}{\sqrt{2}} \right)^{2m}
\times }
\nonumber\\
& & \sum_{ m_1,\cdots,m_{\ell} \geq 0 \atop m_1+ \cdots + m_{\ell}=m }
\left[ \prod_{j=1}^{\ell}
tr_{\mbox{\footnotesize f}}  \left( \sum_{n_j=0}^{2m_j}\frac{1}
 { ({k \llap /}^{(j)})^{2m_j+1-n_j} (p \llap / + {k \llap /}^{(j)})^{n_j+1}}
 \right) \right] \ .
\label{expan}
\end{eqnarray}
This identity can be checked in a straightforward way by expanding the
corresponding fractions on the l.h.s. in powers of
$g H_0/(\sqrt{2} {k \llap /}^{(j)})$
and $g H_0/(\sqrt{2}({p \llap /} + {k \llap /}^{(j)}))$. Here we note that
this identity holds only as long as these two quantities have norms smaller
than one. For small internal momenta $k$ and/or $(p+k)$ (smaller than
$(g H_0/{\sqrt{2}})$)~\footnote
{In the physically interesting region, i.e., near the minimum of
$V_{\mbox{\footnotesize eff}}$,
we have: $(g H_0/\sqrt{2}) \approx m_{t} $. },
i.e., in the region of non-perturbative internal momenta, the above
identity is, strictly speaking, not valid, because the perturbative
``massless'' sum on the r.h.s.~is formally divergent in such a case. However,
the ``massive'' l.h.s.~is finite even for such small momenta, and it
represents in this case the analytic continuation of the r.h.s.
This identity, applied to (\ref{Vl1}), represents the window through which
we get from a perturbative (diagrammatic) approach into the nonperturbative
physics of the dynamical symmetry breaking - through analytic continuation.

Therefore, we can rewrite (\ref{Vl1})
\begin{equation}
V_{\mbox{\footnotesize eff}}^{(\ell+1)-loops}(H_0) = \frac{i}{2\ell}
\int \frac{d^4p}{(2\pi)^4}
 \left[ \int \frac{d^4k}{(2\pi)^4} \frac{i G N_{\mbox{\footnotesize c}}}{2}
 tr_{\mbox{\footnotesize f}}  \left(
\frac{1}{({k \llap /}-\frac{g H_0}{\sqrt{2}})
(p \llap / + {k \llap /}-\frac{g H_0}{\sqrt{2}})} \right) \right]^{\ell}
 \ .
\label{Vl2}
\end{equation}
Performing Wick's rotation into the Euclidean metric ($d^4k \mapsto
i d^4 \bar k$, ${k \llap /} \mapsto {\bar k \llap /}$), we can rewrite
the above integral over $k$ in the Euclidean and cut-off regularized
form
\begin{displaymath}
\int \frac{i d^4 k}{(2 \pi)^4} tr_{\mbox{\footnotesize f}}  \left(
\frac{1}{({k \llap /}-\frac{g H_0}{\sqrt{2}})
(p \llap / + {k \llap /}-\frac{g H_0}{\sqrt{2}})} \right) =
\end{displaymath}
\begin{equation}
= -\int_{{\bar k}^2 \leq \Lambda^2} \frac{d^4 \bar k}{(2 \pi)^4}
tr_{\mbox{\footnotesize f}}  \left[
\frac{1}{({\bar k \llap /}-\frac{g H_0}{\sqrt{2}})
({\bar p \llap /} + {\bar k \llap /}-\frac{g H_0}{\sqrt{2}})} \right]
= 4{\cal{K}}_{H}  ({\bar p}^2, H_0^2) \ .
\label{Ktr}
\end{equation}
Here we used the notation of eq.~(\ref{Aomega}) for ${\cal{K}}_{H} $.
Using further $d^4p \mapsto i d^4 \bar p$
and rescaling ${\bar p}^2 (\leq \Lambda^2)
\mapsto \Lambda^2 {\bar p}^2$ yields the integral
\begin{equation}
V_{\mbox{\footnotesize eff}}^
{(\ell+1)-loops}(H_0) = - \Lambda^4 \int_{{\bar p}^2 \leq 1}
\frac{d^4{\bar p}}{(2\pi)^4} \frac{1}{2\ell} \left[
\frac{G N_{\mbox{\footnotesize c}} \Lambda^2}{8 \pi^2}
{\cal{J}}_{H}  ({\bar p}^2, {\varepsilon}^2) \right]^{\ell} \ ,
\label{Vl3}
\end{equation}
where we employed the notation of eq.~(\ref{Jdef}) for ${\cal{J}}_{H} $ and
${\varepsilon}$.
Summing finally over $\ell = 1,2,\ldots$ yields the logarithmic series and
the result identical to $V^{\mbox{\scriptsize (ntl)}}
_{\mbox{\footnotesize eff}} $ of eq.~(\ref{res2})
\begin{equation}
\sum_{\ell=1}^{\infty} V_{\mbox{\footnotesize eff}}^{(\ell+1)-loops}(H_0) =
 \frac{\Lambda^4}{2 (4\pi)^2} \int_{{\bar p}^2 \leq 1}
d {\bar p}^2 {\bar p}^2 \ln \left[ 1- \frac{(G N_{\mbox{\footnotesize c}}
\Lambda^2)}{(8 \pi^2)}
 {\cal{J}}_{H}  ({\bar p}^2, {\varepsilon}^2) \right] \equiv
 V^{\mbox{\scriptsize (ntl)}} _{\mbox{\footnotesize eff}}  \ .
\label{resC}
\end{equation}
Therefore, we can really interpret the
$V^{\mbox{\scriptsize (ntl)}}_{\mbox{\footnotesize eff}} $ of (\ref{res2})
diagrammatically as the contribution of all those diagrams beyond
the one loop which are next-to-leading order in the formal
$1/N_{\mbox{\footnotesize c}}$-expansion,
provided $(G N_{\mbox{\footnotesize c}} \Lambda^2)/(8 \pi^2) = {\cal{O}} (1)$
as suggested by the 1-loop gap equation (\ref{1lgap}). On the other hand,
the $V^{\mbox{\scriptsize (ntl)}}_{\mbox{\footnotesize eff}} $ in
eq.~(\ref{res2}) was derived there by path integral
method as the contribution of the {\it quadratic} fluctuations
${\sigma}_1({\bar x})^2$ (cf.~eq.~(\ref{omega})) of the Higgs field
around its ``classical'' value $H_0$.

By arguments very similar to those at the beginning of this Appendix,
we can demonstrate that the dominant (in the
$1/N_{\mbox{\footnotesize c}}$-expansion) 1-PI
diagrams contributing to the mass renormalization of $m_{t} $ are those of
Figs.~5a-c - these are all the 1-PI diagrams of order
$1/N_{\mbox{\footnotesize c}}$ with two
top quark outer legs. However, unlike the diagrams in Figs.~2-4, this
time the top quark propagators in the loops have the non-zero {\it bare}
mass $m_{t} $ ($=m_{t} (\Lambda)$) - i.e., the mass obtained
from the gap equation (\ref{2lgap}) (cf.~Table 1, fifth column, or
third column when QCD effects included). These
graphs lead to the corrected propagator
\begin{equation}
D_{t} (q) = \frac{i}{{q \llap /}-m_{t} -{\tilde \pi}_{t} ({q \llap /})
+ i \epsilon} \ , \qquad {\tilde \pi}_{t} ({q \llap /}) =
\sum_{\ell=0}^{\infty} {\tilde \pi}_{t} ^{(\ell+1)}({q \llap /}) \ ,
\qquad m_{t} ^{\mbox{\footnotesize ren.}} = m_{t}
+ {\tilde \pi}({q \llap /})|_{{q \llap /}=m_{t} } \ .
\label{tprop}
\end{equation}
With ${\tilde \pi}_{t} ^{(\ell+1)}({q \llap /})$ we denote the contribution
of the $(\ell+1)$-loop 1-PI diagram of Figs.~5, i.e., the one containing
$\ell$ top quark loops with bare mass $m_{t} $. Using an approach similar
to the one employed earlier in this Appendix for deriving the Green
functions
${\tilde \Gamma}_{H} ^{(2m;\ell+1)}$, and again carefully counting the
number of possible contractions of $H$ and $\bar t t$ (in the formalism
of the Wick theorem), we obtain
\begin{equation}
{\tilde \pi}_{t} ^{(\ell+1)}({\bar {q \llap /}}) =
(-1)\frac{1}{N_{\mbox{\footnotesize c}}}
\left( \frac{G N_{\mbox{\footnotesize c}}}{2} \right) \int_{{\bar p}^2
 \leq \Lambda^2}
 \frac{d^4{\bar p}}{(2 \pi)^4} \lbrack 2 G N_{\mbox{\footnotesize c}}
 {\cal{K}}_{H} ({\bar p}^2, m_{t} ^2)
 \rbrack^{\ell} \frac{({\bar {p \llap /}} + {\bar {q \llap /}} + m_{t} )}{
 \lbrack ({\bar p} + {\bar q})^2 + m_{t} ^2 \rbrack } \ ,
 \label{pitl}
 \end{equation}
 where we used again the Euclidean metric with the simple spherical cut-off,
 and the notation (\ref{Aomega}) for the function ${\cal{K}}_{H} $.
 Summing up all these contributions, rescaling the momentum ${\bar p} \mapsto
 \Lambda {\bar p}$, and using the notations of (\ref{Jdef}) and (\ref{1lgap}),
 we arrive at
 \begin{equation}
 \frac{{\tilde \pi}_{t} ({\bar {q \llap /}})}{\Lambda} =
 \frac{(-1)}{N_{\mbox{\footnotesize c}}} 4 \pi^2 a
 \int_{{\bar p}^2 \leq 1} \frac{d^4{\bar p}}{(2 \pi)^4}
 \frac{1}{\lbrack 1 - a {\cal{J}}_{H} ({\bar p}^2, (m_{t} /\Lambda)^2)
 \rbrack}  \frac{\lbrack {\bar {p \llap /}}
   + ({\bar {q \llap /}}/\Lambda) + (m_{t} /\Lambda) \rbrack}{
 \lbrack ({\bar p} + ({\bar q}/\Lambda))^2 + (m_{t} /\Lambda)^2 \rbrack} \ .
 \label{pit}
\end{equation}
The angular integration in this expression can be performed in a
straightforward way by using 4-dimensional spherical coordinates in
the coordinate system in which ${\bar q} =
(0,0,0,|{\bar q}|)$ ($\Rightarrow {\bar p} \cdot {\bar q}
= |{\bar p}| |{\bar q}| \cos{\theta}$; ${\bar {q \llap /}}
= {\bar \gamma^3} |{\bar q}|$) and the integral (\ref{rgang}).
After performing the angular integrations in
(\ref{pit}), we are left with radial integration alone
\begin{eqnarray}
 \frac{{\tilde \pi}_{t} ({\bar {q \llap /}})}{\Lambda}{\bigg|}_{
{\bar {q \llap /}}=m_{t} } & = &
\frac{1}{4 N_{\mbox{\footnotesize c}}} a \int_0^1 \frac{d |\bar p|}
{\lbrack 1-a{\cal{J}}_{H} (|\bar p|^2,
x^2) \rbrack } {\Big [} \frac{1}{2 x^3} |\bar p|^3 \left( |\bar p|^2 + 2 x^2
- |\bar p| \sqrt{(|\bar p|^2 + 4 x^2)} \right) +
\nonumber\\
& &
+ \frac{2}{x} |\bar p|^2 \left( |\bar p| - \sqrt{(|\bar p|^2 + 4 x^2)}
\right) {\Big ] } \ ,
\label{pitrad}
\end{eqnarray}
where $x=(m_{t} /\Lambda)$, and we took into account the on-shell
condition (${\bar {q \llap /}} = {q \llap /} = m_{t} $, ${\bar q}^2 = -q^2 =
-m_{t} ^2$). This gives us the renormalized mass
$m_{t} ^{\mbox{\footnotesize ren.}}$ in
eq.~(\ref{mtren}), according to (\ref{tprop}) - when we ignore
the QCD effects.

The inclusion of the leading (i.e., 1-loop) QCD effects to the
renormalization of $m_{t} $ can be obtained from the diagram of Fig.~5a,
where the dotted line now represents the gluon. This effect is
well-known in the literature. It results in the following expression
\begin{equation}
(\delta m_{t} )^{\mbox{\scriptsize QCD}}_{\alpha \beta} =
- i \frac{g_{\mbox{\scriptsize s}}^2}{4} \sum_{a,\gamma}
{\lambda}^{a}_{\alpha \gamma}
{\lambda}^{a}_{\gamma \beta} \int \frac{d^4p}{(2\pi)^4} {\Big \{ }
\gamma^{\mu} \frac{i}{\left( {q \llap /}-{p \llap /} - m_{t}  \right) }
\gamma^{\nu} \frac{(-i)}{p^2} \left[ g_{\mu\nu}-\left( 1-\xi \right)
\frac{p_{\mu} p_{\nu}}{p^2} \right] {\Big \} } \ ,
\end{equation}
where $\alpha$ and $\beta$ are the color indices of the top quark,
$\lambda^{a}/2$ are the $SU(3)_c$-generators,
$g_{\mbox{\scriptsize s}}$ is the QCD gauge coupling
($\alpha_{\mbox{\footnotesize s}} = g_{\mbox{\scriptsize s}}^2/(4\pi)$),
and the top quark is taken on-shell (${q \llap /} = m_{t} $). It turns out
that the gauge-dependent part of the gluon propagator (proportional
to $(1-\xi)$) does not contribute to
$(\delta m_{t} )^{\mbox{\scriptsize QCD}}$.
Using the know relation of QCD
\begin{displaymath}
\sum_{a,\gamma} {\lambda}^{a}_{\alpha \gamma}
{\lambda}^{a}_{\gamma \beta} = \frac{2 (N_{\mbox{\footnotesize c}}^2-1)}
{N_{\mbox{\footnotesize c}}} \delta_{\alpha \beta} \ ,
\end{displaymath}
and the well-known techniques of rewriting the denominators as
integrals of exponentials, we end up after some algebraic manipulations
with the following (1-loop) result
\begin{equation}
\left( \delta m_{t}  \right)^{\mbox{\scriptsize QCD}}_{\alpha\beta} =
\delta_{\alpha\beta} \frac{g_{\mbox{\scriptsize s}}^2
(N_{\mbox{\footnotesize c}}^2-1)}{16 \pi^2 N_{\mbox{\footnotesize c}}} m_{t}
\int_0^1 dz (1+z) \int_0^{\infty} \frac{d\tau}{\tau}
\rho_{\mbox{\footnotesize f}}
\left( {\tau,\Lambda_{\mbox{\footnotesize f}} }
\right) \exp \left(-m_{t} ^2 z \tau \right) \ .
\label{dmQCD}
\end{equation}
Here we inserted the usual regulator $\rho_{\mbox{\footnotesize f}}
(\tau,\Lambda_{\mbox{\footnotesize f}} )$ to make the
integral finite. For example, $\rho_{\mbox{\footnotesize f}} $ may
mean that we make a simple Pauli-Villars subtraction (with mass
$\Lambda_{\mbox{\footnotesize f}}$ replacing $m_{t} $).
We can also choose the usual ``proper-time'' cut-off:
$\rho_{\mbox{\footnotesize f}}  = 0$
for $\tau < 1/\Lambda_{\mbox{\footnotesize f}} ^2$, and
$\rho_{\mbox{\footnotesize f}} =1$ otherwise. In the results
of this paper, we have, for simplicity, always equated the cut-offs
for the quark and the bosonic momenta:
$\Lambda_{\mbox{\footnotesize f}}  =$$ \Lambda_{\mbox{\footnotesize b}}  =$
$ \Lambda$.

\end{appendix}

\newpage

\vspace{4cm}

\oddsidemargin-2.7cm
\evensidemargin-2.7cm

\begin{table}[h]
\vspace{0.3cm}

\begin{center}
Table 1 \\
\vspace{0.5cm}
\begin{tabular}{|c|c|c|c|c|c|} \hline
$a= \frac{ N_{\mbox{\scriptsize c}} G \Lambda^2}{8 \pi^2}$ &
$\sqrt{z_1}= \frac{ m_t^{(1)}}{\Lambda}$ &
$\sqrt{z}= \frac{ m_t^{(H+ gl)}}{\Lambda}$ &
$\frac{\left( m_t^{(H+ gl)} \right)
^{\mbox{\scriptsize ren.}}}{\Lambda}$ &
$\frac{ m_t^{(H)} }{\Lambda}$ &
$\frac{ \left( m_t^{(H)}\right)
^{\mbox{\scriptsize ren.}}}{\Lambda}$ \\ \hline
2.198 & 0.70 & 0.669 & 0.575 & 0.657 & 0.532 \\
1.918 & 0.60 & 0.555 & 0.477 & 0.538 & 0.429 \\
1.673 & 0.50 & 0.432 & 0.368 & 0.399 & 0.303 \\
1.605 & 0.469& 0.389 & 0.328 & 0.332 & 0.237  \\
1.5970 & 0.46543& 0.384& 0.323& 0.309 & 0.213 \\
1.5969 & 0.4654& 0.384& 0.323&   -   &   -  \\
1.564 & 0.45 & 0.359 & 0.300 & - & - \\
1.523 & 0.43 & 0.321 & 0.264 & - & - \\
1.506 & 0.4215&0.299 & 0.241 & - & - \\
1.503 & 0.42 & 0.293 & 0.235 & - & - \\
1.4988 & 0.41785&0.277 & 0.217 & - & - \\
1.4987 & 0.4178& - & - & - & - \\ \hline
\end{tabular}
\end{center}
\end{table}

\oddsidemargin-0.5cm
\evensidemargin-0.5cm

\clearpage

\section{Table and figure captions}

\noindent {\bf Table 1}: The solutions $\sqrt{z} = (m_{t} /\Lambda)$
(3rd column) of the improved
gap equation (\ref{2lgap}), for a given input $a=(N_{\mbox{\footnotesize c}}
G \Lambda^2)/(8 \pi^2)$
(1st column), where $G$ is the 4-fermion coupling of the TSM (see (\ref{TSM}))
and $\Lambda$ is the upper energy cut-off (i.e., roughly the energy at which
the condensation takes place). The last entry in a column
corresponds to the disappearance of the non-trivial minimum.
The 4th column contains the corresponding
renormalized top quark masses. The 5th and the 6th columns contain
the solutions $(m_{t} /\Lambda)$ and their renormalized values, respectively,
for the case when the QCD effects are ignored.
$N_{\mbox{\footnotesize c}} = 3$ was taken.

\vspace{1cm}

\noindent {\bf Figs.~1(a)-1(e)}: Effective potential
$V_{\mbox{\footnotesize eff}}$ as function of
${\varepsilon}^2= H_0^2 (G M_0^2) / (2 \Lambda^2)$. For convenience,
$V_{\mbox{\footnotesize eff}}$ is multiplied by
$\chi = 8 \pi^2/(N_{\mbox{\footnotesize c}} \Lambda^4)$.
The notation ``ntl(H+gluons)'' means that
all the next-to-leading $1/N_{\mbox{\footnotesize c}}$-contributions of the
Higgs sector and the dominant part of the QCD contributions are included;
``ntl(H)'' means the same, but without QCD. Where possible, we
also included the curves for the case of (0+1+2)-loop (without QCD)
and (0+1)-loop calculation.
We took $N_{\mbox{\footnotesize c}}=3$ and the input values
$a=(N_{\mbox{\footnotesize c}} G \Lambda^2)/(8\pi^2)
\simeq 2.198$, $1.673$, $1.564$, $1.5031$, $1.4982$, respectively
(corresponding to the 1-loop minima $\sqrt{z_1}= m_{t} ^{(1)}/\Lambda = 0.7$,
$0.5$, $0.45$, $0.42$, $0.4175$, respectively).

\vspace{1cm}

\noindent {\bf Figs.~2(a)-(c)}: The 1-loop 1-PI diagrams contributing to
1-PI Green functions $\tilde \Gamma_{H} ^{(2m; 1)}(p_1, \ldots, p_{2m})$;
incidentally, they give at the same time the leading order terms in the formal
$1/N_{\mbox{\footnotesize c}}$-expansion of $V_{\mbox{\footnotesize eff}}$
(of order ${\cal{O}} (N_{\mbox{\footnotesize c}})$). Full lines
represent massless top quarks, and dotted lines the scalar non-dynamical
Higgs of the Lagrangian (\ref{TSM1}) (same notation also in Figs.~4, 5, 6).

\vspace{1cm}

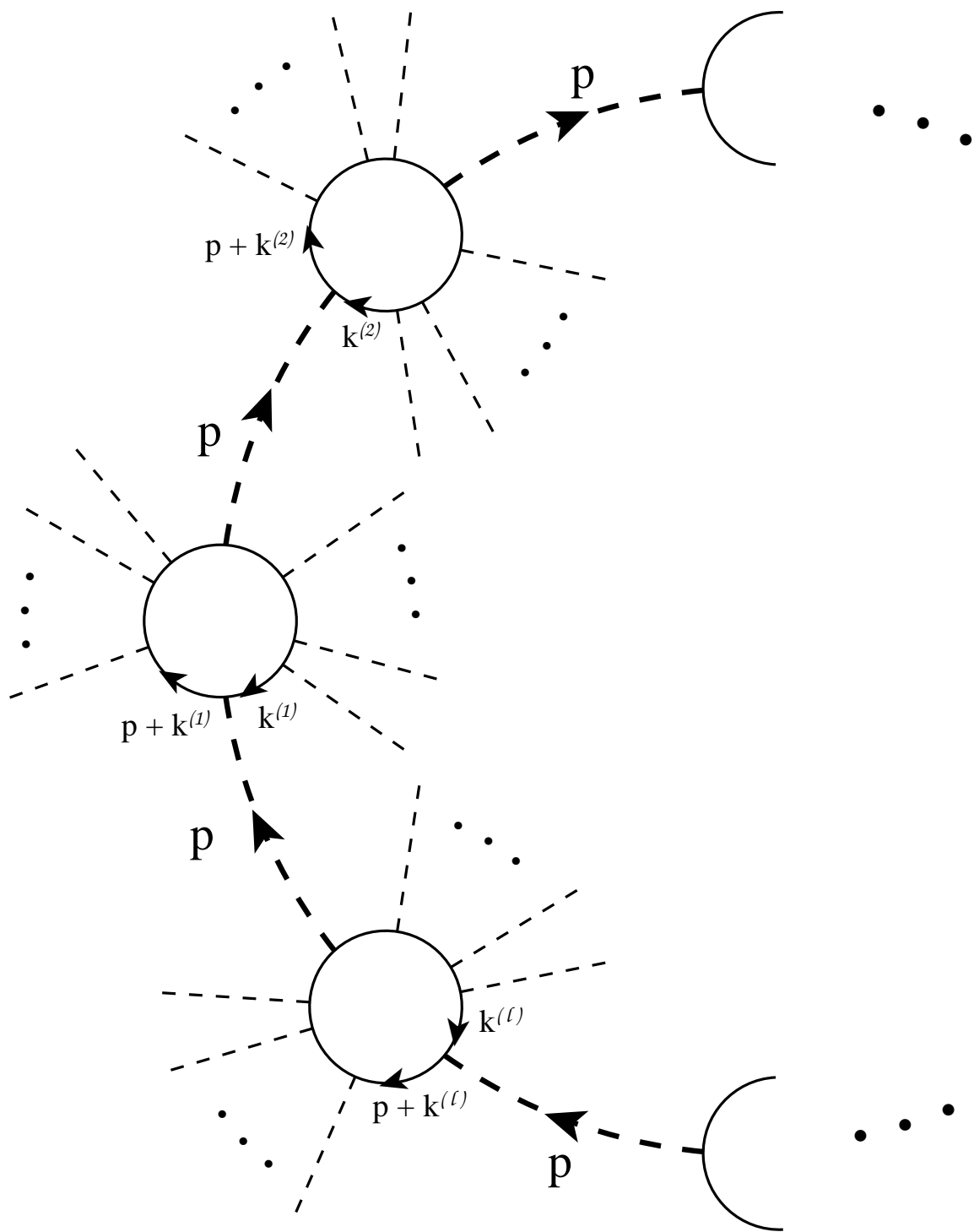
\noindent {\bf Fig.~3}: The $(\ell + 1)$-loop 1-PI diagrams which contribute
to the 1-PI Green functions $\tilde \Gamma_{H} ^{(2m; \ell+1)}(p_1, \ldots,
p_{2m})\vert_{\lbrace p_k \rbrace = \lbrace 0 \rbrace}$
and which yield the next-to-leading order terms (beyond 1-loop) in the
formal $1/N_{\mbox{\footnotesize c}}$-expansion of
$V_{\mbox{\footnotesize eff}}$ (of order ${\cal{O}}
((G N_{\mbox{\footnotesize c}})^{\ell}) = {\cal{O}} (1)$).
The diagrams contain $\ell$ loops of (massless) top quarks.
These loops are connected into another circle by $\ell$ propagators
of the (non-dynamical) Higgs. The $j^{th}$ fermionic loop ($j=1, \ldots,
\ell$) has $(2m_j)$ outer legs of Higgses with zero momenta attached
to it ($n_j$ and $(2m_j-n_j)$ on the ``outer'' and the ``inner''
halves, respectively). The total number of outer legs with zero
momenta is $2(m_1 + \cdots + m_{\ell}) = 2m$.

\vspace{1cm}

\noindent {\bf Figs.~4(a)-(b)}: Examples of 3-loop and 4-loop 1-PI diagrams
which,
while in principle contributing to $\tilde \Gamma_{H} ^{(2m; 3)}(p_1, \ldots,
p_{2m})$, do not contribute to the next-to-leading order
terms (${\cal{O}} (1)$)
of $V_{\mbox{\footnotesize eff}}$ in the formal
$1/N_{\mbox{\footnotesize c}}$-expansion, but rather to terms
${\cal{O}}(G^2 N_{\mbox{\footnotesize c}})=
{\cal{O}}(G^3 N_{\mbox{\footnotesize c}}^2)=
{\cal{O}}(1/N_{\mbox{\footnotesize c}})$. This is so
because the colors
of all the (massless) top quarks in a given quark loop are forced to be
equal and the number of quark loops is smaller than that in the 3-loop
and 4-loop diagrams of Fig.~3, respectively.

\vspace{1cm}

\noindent {\bf Figs.~5(a)-(c)}: the 1-PI diagrams with two outer top quark
legs which give the leading (${\cal{O}}(1/N_{\mbox{\footnotesize c}})$)
contribution to the
renormalization of the mass $m_{t} $. Unlike the diagrams of Figs.~2-4,
the top quark propagators here contain the non-zero bare mass $m_{t} $
which was the solution to the gap equation in the next-to-leading
order in $1/N_{\mbox{\footnotesize c}}$.

\end{document}